\documentclass{aa}

\usepackage{graphicx}
\usepackage{txfonts}
\usepackage[colorlinks=true]{hyperref}
\usepackage{xcolor,colortbl}
\usepackage{float}
\usepackage{adjustbox}
\usepackage{pdflscape}
\usepackage{lscape}
\usepackage{comment}
\usepackage{changepage}
\usepackage{geometry}
\usepackage{threeparttable}
\hypersetup{colorlinks=true,
    linkcolor=blue,
    urlcolor=blue,
    citecolor=blue}

\usepackage{makecell}
\makeatletter
\renewcommand*\aa@pageof{, page \thepage{} of \pageref*{LastPage}}
\makeatother
\usepackage{diagbox}
\usepackage{multirow}
\usepackage{graphicx} %package to manage images
\graphicspath{ {./images/} }

\usepackage[rightcaption]{sidecap}

\usepackage{wrapfig}
\usepackage{tabularx}
\usepackage[caption=false]{subfig}
\begin{document} 
 
   \title{From seagull to hummingbird: New diagnostic methods for resolving galaxy activity}
   % \subtitle{From seagull to hummingbird}

   % FROM SEAGULL TO HUMMINGBIRD

%   \subtitle{I. Overviewing the $\kappa$-mechanism}

   \author{C. Daoutis \inst{1,2,3}
          \and
          A. Zezas \inst{1,2,3}
          \and
          E. Kyritsis\inst{1,2}
          \and 
          K. Kouroumpatzakis \inst{4}
          \and
          P. Bonfini \inst{1,5,6}
          }

   \institute{Physics Department, and Institute of Theoretical and Computational Physics, University of Crete, 71003 Heraklion, Greece \\
         \email{cdaoutis@physics.uoc.gr}
         \and
             Institute of Astrophysics, Foundation for Research and Technology-Hellas, 71110 Heraklion, Greece
        \and
            Center for Astrophysics | Harvard \& Smithsonian, 60 Garden St., Cambridge, MA 02138, USA
        \and 
            Astronomical Institute, Academy of Sciences, Bo\v{c}n\'{\i} II 1401, CZ-14131 Prague, Czech Republic
        \and
            ALMA Sistemi Srl, Guidonia (Rome), 00012, Italy
        \and
            Quantum Innovation Pc, Chania, 73100, Greece}

   \date{Received 1 July 2024 / Accepted: 7 November 2024}

% \abstract{}{}{}{}{} 
% 5 {} token are mandatory
 
  \abstract
   {One of the principal challenges in astrophysics involves the classification of galaxies based on their activity. Presently, the characterization of galactic activity usually requires multiple diagnostics to fully cover the diverse spectrum of galaxy activity types. Additionally, the presence of multiple sources of excitation with similar observational signatures hinders the exploration of the activity of a galaxy.} %leave it empty if necessary  
   {In this study, our objective is to develop an activity diagnostic tool that addresses the degeneracy inherent in the existing
   emission line diagnostics by identifying the underlying excitation mechanisms of a mixed-activity galaxy 
   principal components (star formation, active nucleus, or old stellar populations) and identifying the dominant ones.}
   {We utilize the random forest machine-learning algorithm, trained on three primary activity classes of star-forming, active galactic nucleus (AGN), and passive, that represent the three key gas excitation mechanisms. This diagnostic relies on four discriminating features: the equivalent widths of three spectral lines--[\ion{O}{III}] $\lambda$5007, [\ion{N}{II}] $\lambda$6584, and H$\alpha$--along with the D4000 continuum break index.}
   {We find that this classifier achieves almost perfect performance scores in the principal activity classes. In particular, the achieved overall accuracy is $\sim$99\%, while the recall scores are: $\sim$100\% for star-forming, $\sim$98\% for AGN, and $\sim$99\% for passive. The nearly perfect scores achieved enable the decomposition of mixed activity classes into the three primary gas excitation mechanisms with high confidence, thereby resolving the degeneracy inherent in current activity classification methods. Furthermore, we find that our classifier scheme can be simplified to a two-dimensional diagnostic diagram of D4000 index against log$_{10}$(EW([\ion{O}{III}])$^{2})$ line without significant loss of its diagnostic power.}
   {We introduce a diagnostic capable of classifying galaxies based on their primary gas excitation mechanisms. Simultaneously, it can deconstruct the activity of mixed-activity galaxies into these principal components. This diagnostic encompasses the entire range of galaxy activity. Additionally, D4000 index serves as a valuable indicator for resolving the degeneracy among various activity components by estimating the age of the stellar populations within a galaxy.}


   \maketitle
%
%-------------------------------------------------------------------

\section{Introduction}

One of the most challenging and important topics in astrophysics is the activity classification of galaxies. Galaxies can be separated into categories based on their prevailing gas excitation mechanism. Typically, the identification of the radiation source that excites the gas is determined from the observed spectrum of a galaxy \citep[e.g., atomic emission lines;][]{1981PASP...93....5B}. 

In a galaxy, three primary gas excitation mechanisms can occur. The first involves the presence of young stars that emit substantial amounts of intense UV radiation, thereby ionizing the surrounding gas cloud. Another distinct mechanism arises from an active black hole situated at the galaxy's center, where circumnuclear material undergoes accretion and produces intense UV and X-ray radiation. Finally, hot-evolved stellar populations can generate sufficient UV radiation to excite the surrounding gas in their vicinity. Furthermore, the spectrum of a galaxy cannot be exclusively linked to a specific excitation mechanism. For instance, a population of post-Asymptotic Giant Branch (post-AGB) stars can emulate an active galaxy \citep{2008MNRAS.391L..29S}. This intricate nature of galaxies complicates the characterization of their activity.

Moreover, the development of highly effective activity diagnostic tools, in terms of completeness and reliability, are critical for demographic surveys of galactic activity. These are a key element for understanding the interplay between interstellar medium, star formation, stellar populations, and AGN activity in galactic cores and hence the general galactic evolution process \citep[e.g.,][]{1997ApJ...487..568H}. 

Numerous approaches have been taken in this direction, leading to the creation of a multitude of diagnostic tools, in an endeavor to address this problem, which is both highly important and complex. Some of these approaches rely on atomic emission-line fluxes \citep[e.g., ratios of Balmer hydrogen lines with forbidden emission lines;][]{1981PASP...93....5B}, while others use infrared colors \citep[e.g.,][]{2012ApJ...748..142D,2012MNRAS.426.3271M,2012ApJ...753...30S,2023A&A...679A..76D}, aiming to pinpoint the primary source of radiation driving the observed galaxy spectrum and thus categorize its activity class. 

One of the most successful and widely used diagnostics is the \cite{1981PASP...93....5B} diagram (hereafter referred to as the BPT diagram). It’s most commonly used version employs the ratios of the first two Balmer series lines (H$\alpha$ and H$\beta$) and two forbidden emission lines ([\ion{O}{III}] $\lambda5007$ and [\ion{N}{II}] $\lambda6584$) to construct a two-dimensional plot of [\ion{O}{III}] $\lambda5007$/H$\beta$ versus [\ion{N}{II}] $\lambda6584$/H$\alpha$. Based on their position on this diagram, galaxies are categorized as active galactic nuclei (AGN), star-forming (SF), or transition objects \citep[TO or composite galaxies, e.g.,][]{1997ApJ...487..568H}. The galaxies on this diagram form a seagull-like shape, with star-forming galaxies situated in the left wing, composite galaxies comprising its body, and AGN positioned in the right wing. Similarly, additional diagrams involving the [\ion{S}{II}] $\lambda \lambda6717,6731$/H$\alpha$ and [\ion{O}{I}] $\lambda6300$/H$\alpha$ ratios have been utilized, expanding the usefulness of emission-line ratios to objects hosting low-ionization sources. As a result, a new activity class was introduced: the class of LINER \citep[Low-Ionization Nuclear Emission-line Region;][]{1980A&A....87..142H}.

Another activity diagnostic tool for galaxies in the optical spectrum has been introduced by \cite{2010MNRAS.403.1036C}. This approach, which can be viewed as a modified BPT diagram, employs optical emission lines and the H$\alpha$ equivalent width to classify galaxies. An advantage of this diagram is its versatility beyond emission spectra allowing the inclusion of objects with weaker emission lines.

Machine learning algorithms are powerful tools for discerning intricate relationships and patterns within data. They have already been applied to numerous problems across various scientific domains. In astrophysics, specifically, examples of their application ranges from stellar classification tasks \citep[e.g.,][]{2022A&A...657A..62K} to galaxy classification challenges, such as classification based on the BPT diagram \citep{2019MNRAS.485.1085S}, the identification of AGN properties \citep[e.g.,][]{2021IAUS..356..335P}, and the classification of galaxy morphology \citep[e.g.,][]{2018MNRAS.476.3661D}.

Up to this point, we have described the activity of galaxies that are exclusively characterized by a singular gas excitation mechanism. However, a galaxy spectrum can rarely be a result of only one gas excitation mechanism. Composite and LINER are two galaxy classes introduced as distinct classes but in practice they are characterized by different sources of excitation or a combination of excitation mechanisms. Identifying the activity source in a composite galaxy can be intricate. Positioned between star-forming and AGN galaxies in the BPT diagram, the spectra of these galaxies can be excited by AGN or young stellar populations. Moreover, recent investigations propose that not all composite galaxies host an active nucleus; instead, the added ionization source may originate from populations of hot-evolved stars \citep{2017ApJ...840...44B,2019AJ....158....2B}. 

The mechanism driving the activity of LINER galaxies is more challenging to interpret. For several years, it was believed that their activity resulted from a low-luminosity active nucleus \citep{1997ApJ...487..568H}. Some studies introduced the notion that the activity can also be attributed to post-AGB stars \citep{1994A&A...292...13B,2008MNRAS.391L..29S,2013A&A...555L...1P}. \cite{2021ApJ...922..156A} provide further evidence that LINERs can be separated into two subclasses based on the hardness of their ionization field. To date, all available diagnostic methods converge on the conclusion that these two classes result from mixed activities, yet they fail to provide detailed insight into the characterization of the true underlying excitation mechanism.

While some of the aforementioned classification methods have demonstrated success in discerning the AGN from star formation based on the hardness of the radiation source, the majority possess limited applicability, complex implementation, or inability to simultaneously include all activity classes. However, none of these methods can effectively characterize complex instances of galaxy activity where combinations of coexisting principal activity mechanisms result in degeneracy in the observed spectrum (e.g., composite galaxies). In this study, we intend to develop a machine-learning diagnostic tool that employs four key features and can effectively categorize galaxies into activity classes based on their similarities (i.e., shared properties) with the three primary excitation mechanisms: star formation, AGN, and emission from hot-evolved stellar populations. To achieve this, we employ a machine-learning approach. Our choice of discriminative features includes the equivalent widths (EW) of [\ion{O}{III}] $\lambda 5007$, [\ion{N}{II}] $\lambda 6584$, H$\alpha$, and the D4000 continuum break index \citep{1999ApJ...527...54B}. Using equivalent widths instead of actual flux values for spectral lines enables us to encompass passive galaxies within a unified classification framework. Additionally, we anticipate that the D4000 index will help to identify the excitation resulting from hot-evolved stellar populations, which is frequently misidentified as emission from active galaxies. This confusion is especially pronounced in cases of mixed activity classes (e.g., composite galaxies). Our ultimate objectives encompass the identification of the dominant gas excitation mechanism in a galaxy, as well as the recognition of combinations of gas excitation mechanisms coexisting within galaxies exhibiting mixed activity classes.

This paper is organized as follows. In Sect. \ref{datasample} we introduce the data sample, selection method of the galaxy activity classes, and data processing. In Sect. \ref{sec3} we present the algorithm used for the development of our diagnostic tool, its training process, and the metrics used to evaluate its performance. In Sect. \ref{res} we present the results and how we treat mixed-activity galaxy classes. In the same section, we introduce a new activity diagnostic diagram tailored to handle all galaxy activity types (active to passive) offering more resolution (in terms of activity) for mixed-activity classes that are not addressed in the current diagnostics. In Sect. \ref{disc} we discuss potential limitations and we compare our diagnostic with other diagnostic methods. In Sect. \ref{concl} we summarize our conclusions.

%--------------------------------------------------------------------
\section{Data sample} \label{datasample}
\subsection{Data acquisition} \label{sec21}

Our data sample consists of a combination of data from two sky surveys. We begin with the MPA-JHU \citep{2003MNRAS.346.1055K,2004MNRAS.351.1151B,2004ApJ...613..898T} DR8 release of the Sloan Digital Sky Survey (SDSS). In this respect we corss-match the galSpecInfo, galSpecIndx, and galSpecLine catalogs. We focus on the equivalent widths (EW) corresponding to H$\alpha$, the doubly ionized forbidden line of oxygen ([\ion{O}{III}] $\lambda$5007), and the singly ionized forbidden line of nitrogen ([\ion{N}{II}] $\lambda$6584). All these EW values are derived from continuum-subtracted spectra, where negative values indicate emission. We also include the D4000 index from the galSpecIndx catalog. This is the continuum break at the 4000 $\AA$ as defined by \cite{1999ApJ...527...54B}. For all emission lines we apply corrections on the line measurements as reported in the process mentioned in \cite{2019MNRAS.485.1085S}.

To identify a set of inactive galaxies, we also rely on ultraviolet photometry data from the Galaxy Evolution Explorer \citep[GALEX;][]{2005ApJ...619L...1M} survey. To achieve this, we cross-match the SDSS sample and the GALEX-SDSS-WISE Legacy Catalog (GSWLC) described by \cite{2016ApJS..227....2S}, using a search radius of 1$^{\prime \prime}$.

Upon aggregating all available data based on the features of interest for each object, the resulting catalog encompasses a total of 206476 galaxies. Next we impose the following quality criteria. Within the MPA-JHU dataset, objects flagged with \texttt{RELIABLE}=0 in the SDSS catalog are omitted from our analysis. For the remaining sample we require signal-to-noise ratio (S/N) > 5 in the continuum around all the considered spectral lines, namely, H$\alpha$, [\ion{O}{III}], and [\ion{N}{II}]. To ensure the reliability of D4000 measurements, we utilize the H$\gamma$ continuum, selecting galaxies with a continuum exhibiting S/N > 5. The rationale for employing H$\gamma$ in lies in its proximity to the 4000 $\AA$, making it a suitable proxy for assessing the quality of the blue continuum.

Moreover, we exclude galaxies with D4000 values set to 0, as our visual inspection revealed erroneous measurements in spectra due to incomplete coverage of the region of the spectrum needed for the calculation of the D4000. Following the application of the aforementioned quality criteria, the resulting galaxy sample comprises 180436 galaxies.

\subsection{Multi-dimensional emission-line classification of active galaxies} \label{sec:22}

For the formulation of our diagnostic, we opted to train a supervised machine learning algorithm, which requires accurate labels for the galaxies that we intend to incorporate in the training process. To obtain these labels (classifications) for our galaxy sample, we employ the Soft Data-Driven Analysis (SoDDA) classifier of \cite{2019MNRAS.485.1085S}, a four-dimensional diagnostic based on four distinct emission-line ratios of log$_{10}$([\ion{N}{II}]/H$\alpha$), log$_{10}$([\ion{S}{II}]/H$\alpha$), log$_{10}$([\ion{O}{I}]/H$\alpha$), and log$_{10}$([\ion{O}{III}]/H$\beta$). This model was formulated by fitting multivariate Gaussian distributions within this four-dimensional emission-line ratio space. This method offers a notable advantage as it concurrently considers all four important features, as opposed to their two-dimensional projection, as observed in the diagnostic introduced by \cite{2006MNRAS.372..961K}. By doing so, we optimize the credibility of the classification outcome while avoiding contradictory classifications. It also gives the probability for a galaxy to belong to each class of four activity classes (star-forming, AGN, LINER, and composite). The class with the highest predicted probability is adopted as the activity class of each galaxy.

The SoDDA employed here is a probabilistic classifier, where the class label for each galaxy is assigned based on the highest predicted probability. To ensure high-confidence classification of every active galaxy (both star-forming and AGN), an additional cut is applied based on the predicted probabilities for each galaxy to belong to all available classes. Thus, we selected galaxies that have probability differences between the first and second predicted classes of 25\%.

\subsection{Classification of passive galaxies} \label{sec:23}

\begin{figure}[h]
\begin{center}
\includegraphics[scale=0.4]{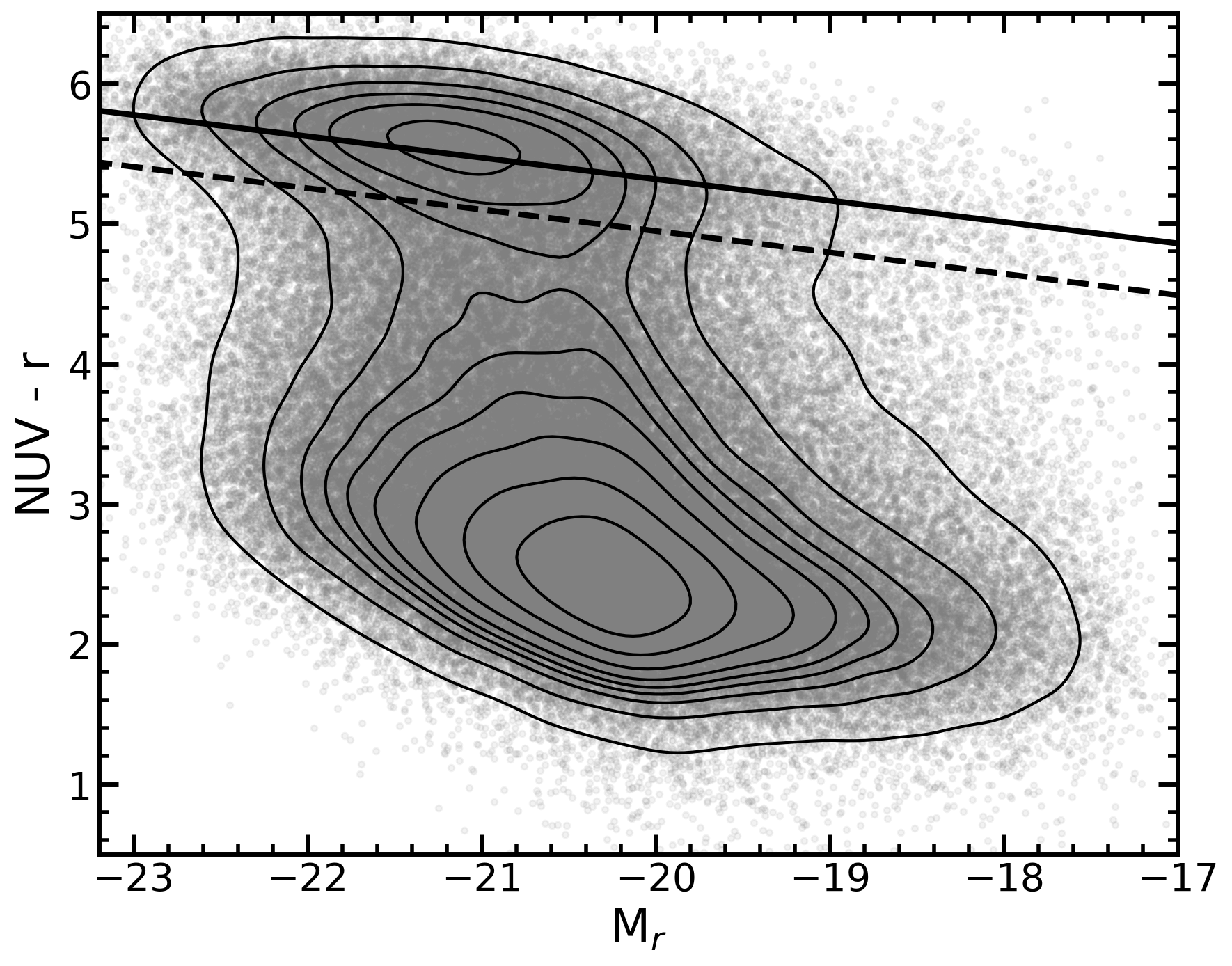}
\end{center}
\caption{CMD diagram of $NUV-r$ against $M_{r}$. The grey dots and black contours represent the entire galaxy sample from which we will select our sample of passive galaxies. The sample of galaxies is taken from the SDSS. The black solid line is the best-fit color-magnitude relation defined by \cite{2008MNRAS.385.1201H}, and the black dashed line is parallel to the former line but displaced by a distance of 1$\sigma$.}
\label{fig:figCMDur}
\end{figure}

After establishing a sample of active galaxies, we also need to define a sample of passive galaxies. For the definition of the initial sample of passive galaxies, we selected galaxies in the red sequence of the color-magnitude diagram \citep[CMD; e.g.,][]{2004ApJ...608..752B,2008MNRAS.385.1201H} of galaxies. The NUV and SDSS r-band photometries are based on the GSWLC and SDSS catalogues respectively. More specifically we use the red sequence definition of \cite{2008MNRAS.385.1201H} on the $NUV - r$ CMD: $NUV$-$r > 5.393 - 0.1782~(M_{r}+20) - 0.370$. In Fig. \ref{fig:figCMDur}, we show this selection criterion by plotting it in a $NUV-r$ against $M_{r}$ CMD, alongside all the galaxies in our sample. To ensure the robustness of the classification, we set S/N > 3 in the $NUV-r$ color for all passive galaxies.

\begin{table}[h]
\caption{Composition of the final sample per galaxy class.} % title of Table
\centering % used for centering table
\begin{tabular}{l c c}
\hline\hline
  Class & Number of objects & Percentage (\%)\\
\hline
Star-forming & 36259 & 61.8 \\
AGN & 1332 & 2.3 \\
Passive & 21090 & 36.0 \\
%[1ex] 
\hline
Total & 58681 & 100.0 \\
\hline
\end{tabular}
\label{table:tab_fs} 
\end{table}

\begin{figure*}[h]
\begin{center}
\includegraphics[scale=0.36]{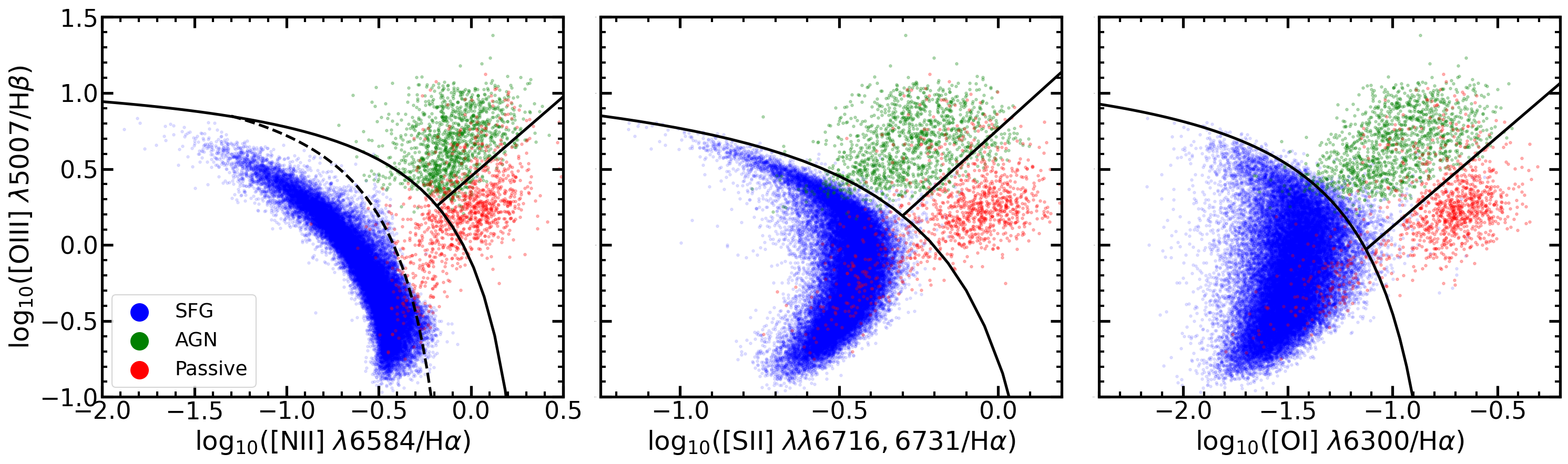}
\end{center}
\caption{Projections of the training sample for the three principal galaxy activity classes. The left plot shows the standard BPT diagram, the middle plot displays log$_{10}$([\ion{O}{III}]/H$\beta$) against log$_{10}$([\ion{S}{II}]/H$\alpha$), and the right plot illustrates log$_{10}$([\ion{O}{III}]/H$\beta$) against log$_{10}$([\ion{O}{I}]/H$\alpha$). The blue dots represent SF galaxies, the green dots represent AGN galaxies, and the red dots represent passive galaxies. Note that in all three plots, only a subset of the passive galaxies is presented due to their typically poor quality of spectra. For these plots, we selected galaxies with emission lines having S/N > 3. In all three plots, the black dashed line corresponds to \cite{2003MNRAS.346.1055K}, the black solid curved line represents \cite{2001ApJ...556..121K}, and the straight black line is \cite{2007MNRAS.382.1415S}, which separates LINERs from AGN.}
\label{fig:figBPT_EL}
\end{figure*}

\begin{figure*}[h]
\begin{center}
\includegraphics[scale=0.48]{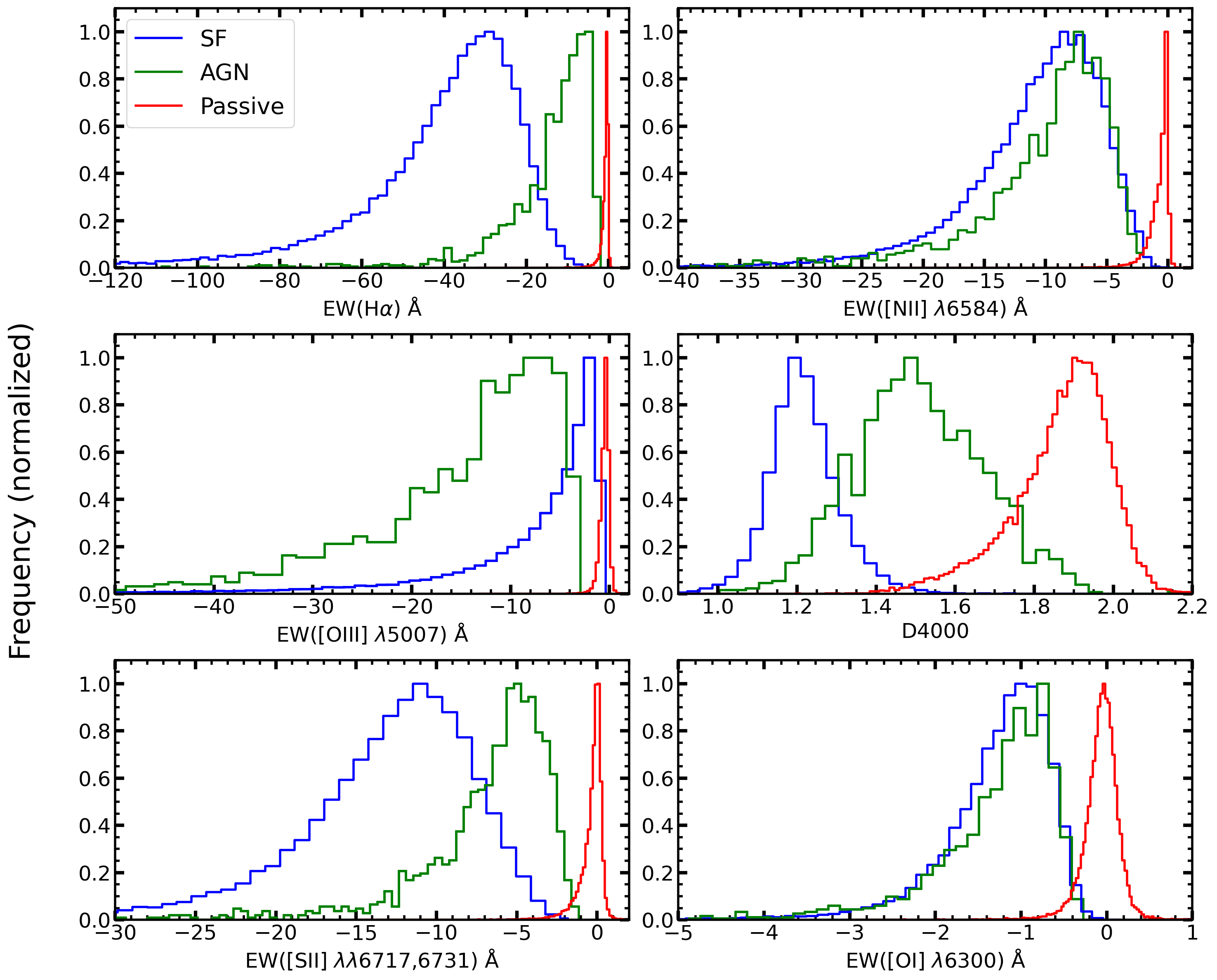}
\end{center}
\caption{Distributions of six potential features for the three principal activity classes: star-forming (SF), AGN, and passive galaxies. In the top left, we observe EW of H$\alpha$; in the top right, EW of [\ion{N}{II}] $\lambda$6584; in the middle left, EW of [\ion{O}{III}] $\lambda$5007; in the middle right, D4000; in the bottom left, EW of [\ion{S}{II}] $\lambda \lambda$6717,6731; and in the bottom right, EW of [\ion{O}{I}] $\lambda$6300. These features represent the equivalent widths of the corresponding emission lines commonly utilized in galactic activity classification models.}
\label{fig:feature_dist}
\end{figure*}

% \begin{table}[h]
% \caption{Report of performance scores calculated on the test sample for each galaxy class using three different metrics.}
% \centering
% \begin{tabular}{l c c c c }
% \hline\hline
% Class & Precision & Recall & F$_{1}$-score & Galaxies \\
% \hline
% Star-forming & 1.00 & 1.00 & 1.00 & 10822\\
% AGN & 0.84 & 0.98 & 0.91 & 384\\
% Passive & 1.00 & 0.99 & 0.99 & 6399 \\
% \hline
% \end{tabular}
% \label{tab:tabpsc}
% \end{table}

\begin{figure}[h]
\begin{center}
\includegraphics[scale=0.4]{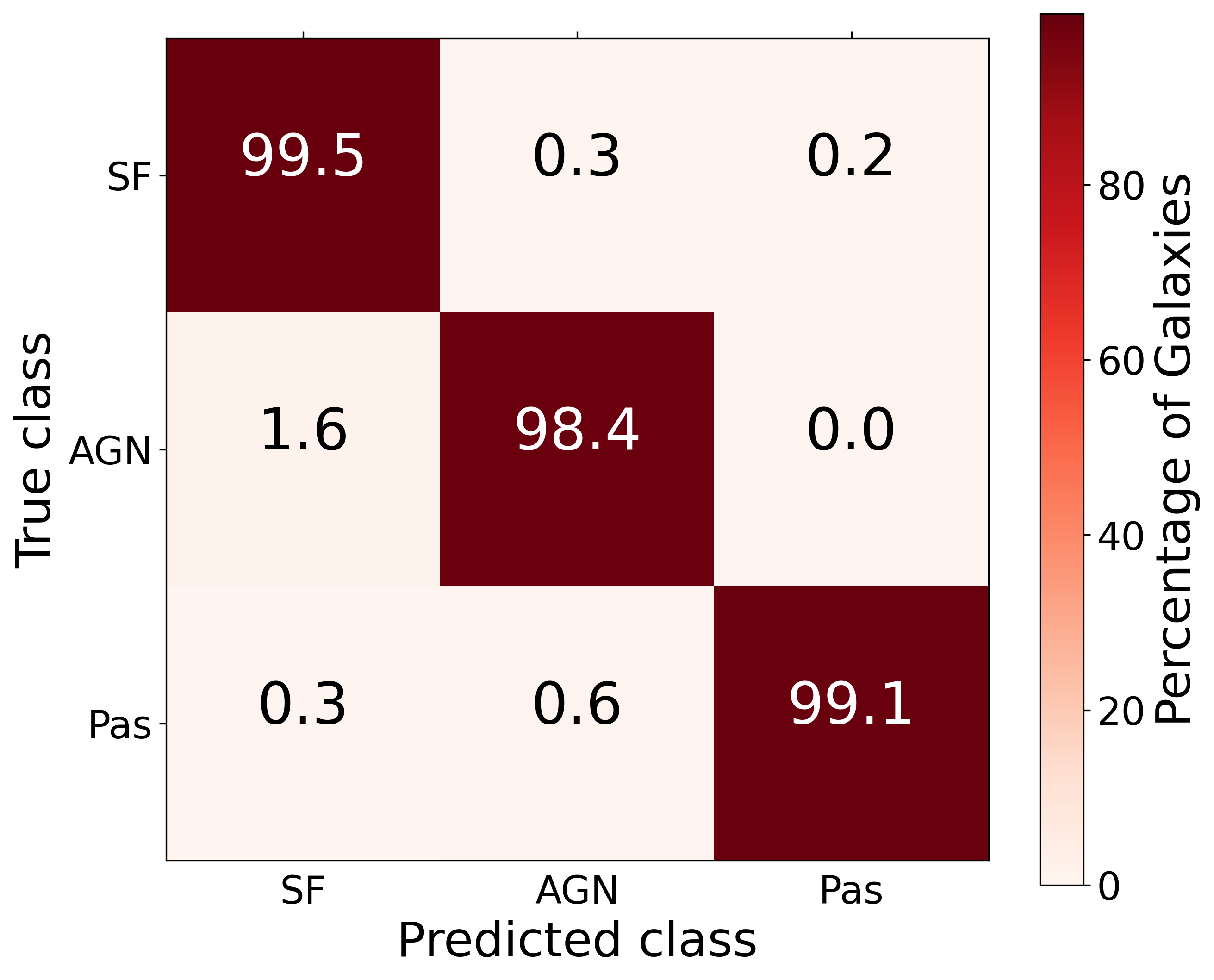}
\end{center}
\caption{Confusion matrix calculated on the test subset of the final sample. The color and the number in each box refer to the percentage of objects calculated relative to the total number of true instances for each class separately. The labels on the x- and y-axis indicate the predicted and true class of a galaxy, respectively. We notice that this matrix is almost diagonal, indicating a high-confidence classification.}
\label{fig:figurercn}
\end{figure}

\begin{figure}[h]
\begin{center}
\includegraphics[scale=0.45]{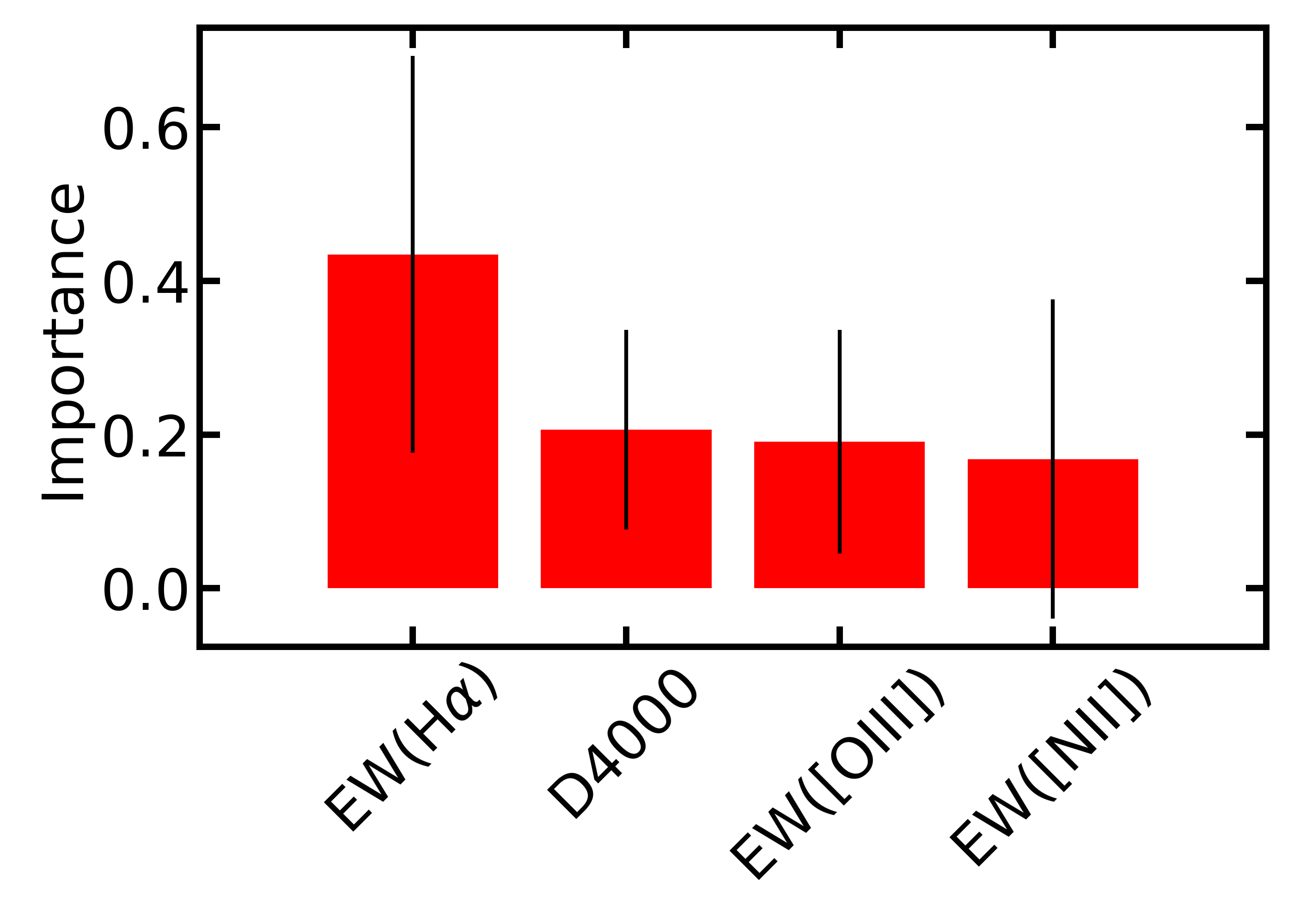}
\end{center}
\caption{Plot of the feature importance calculated during the training process of the algorithm. Error bars are indicative of the standard deviation, providing insight into the stability of the feature importance estimates. The feature with the highest importance score in our diagnostic is EW(H$\alpha$), while all other features exhibit relatively similar levels of relevance.}
\label{fig:ftip}
\end{figure}

% \begin{figure*}[h]
% \begin{center}
% \includegraphics[scale=0.63]{plots/dec_comp_liner_gen.png}
% \end{center}
% \caption{Two BPT plots of log$_{10}$([\ion{O}{III}] $\lambda$5007/H$\beta$) against log$_{10}$([\ion{N}{II}] $\lambda$6584/H$\alpha$) displaying the outcome of the activity decomposition on spectroscopically selected sample of composite (left) and LINER (right) galaxies. Following the application of our diagnostic to these samples, we employ a color-coding scheme based on the highest likelihood of similarity to one of the principal activity classes. On the left, blue contours represent composites predicted as SF (SF-composite), green as AGN (AGN-composite), and red as passive (passive-composite). This plot reveals the overlap of some SF and passive predicted composites. On the right, green contours represent LINERs that have been predicted as AGN and red as passive. This plot shows that there is some overlap between the LINERs predicted as AGN and as passive. Black boundary lines are the same as defined in Fig. \ref{fig:figBPT_EL}.}
% \label{fig:merged_decomp_liner}
% \end{figure*}

 \begin{figure*}
\sidecaption
\includegraphics[width=12cm,height=7cm]{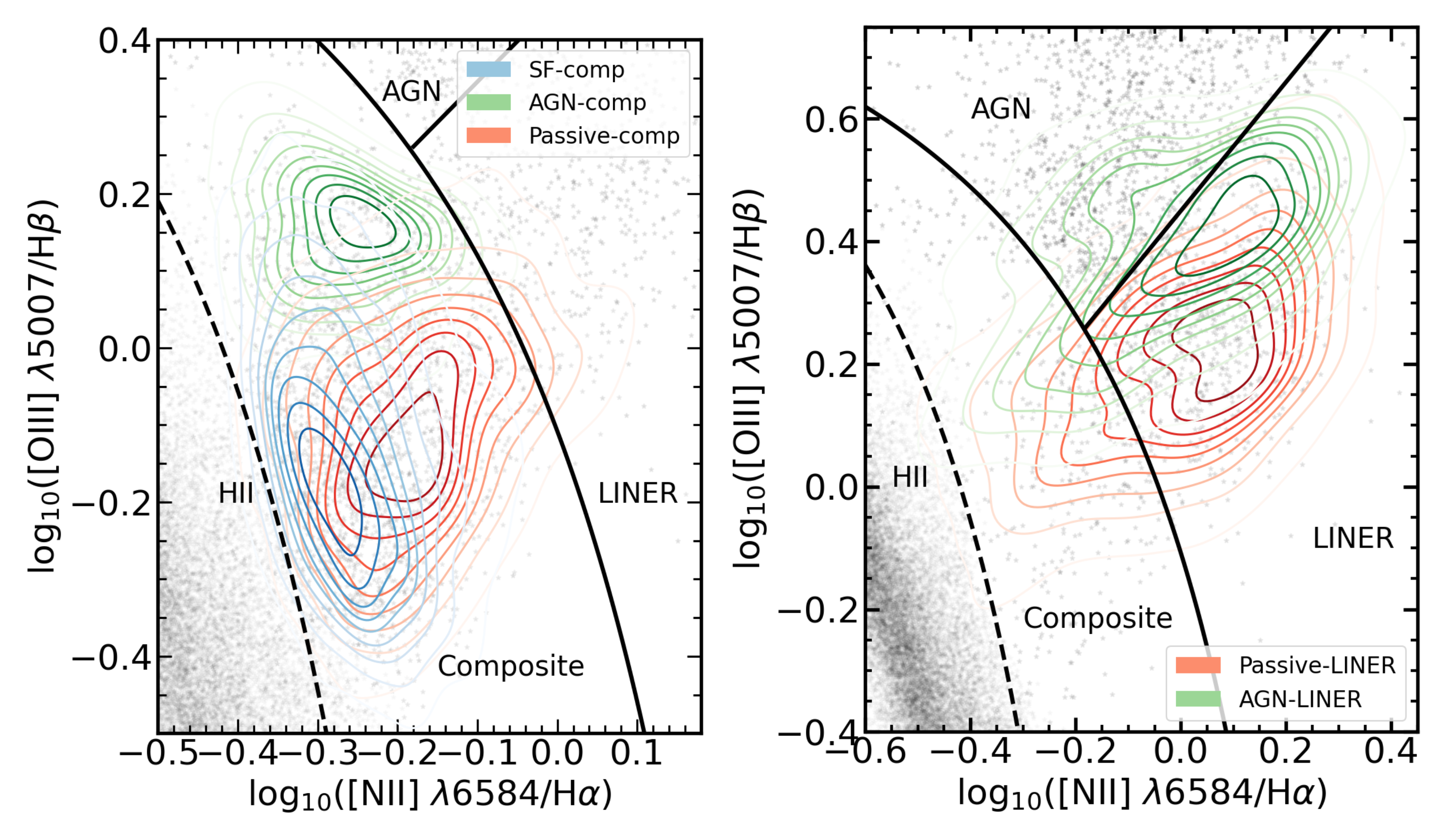}
     \caption{Two BPT plots of log$_{10}$([\ion{O}{III}] $\lambda$5007/H$\beta$) against log$_{10}$([\ion{N}{II}] $\lambda$6584/H$\alpha$) displaying the outcome of the activity decomposition on spectroscopically selected sample of composite (left) and LINER (right) galaxies. Following the application of our diagnostic to these samples, we employ a color-coding scheme based on the highest likelihood of similarity to one of the principal activity classes. On the left, blue contours represent composites predicted as SF (SF-composite), green as AGN (AGN-composite), and red as passive (passive-composite). This plot reveals the overlap of some SF and passive predicted composites. On the right, green contours represent LINERs that have been predicted as AGN and red as passive. This plot shows that there is some overlap between the LINERs predicted as AGN and as passive. Black boundary lines are the same as defined in Fig. \ref{fig:figBPT_EL}.}
    \label{fig:merged_decomp_liner}
\end{figure*}

\begin{figure}[h]
\begin{center}
\includegraphics[width=9cm,height=13cm]{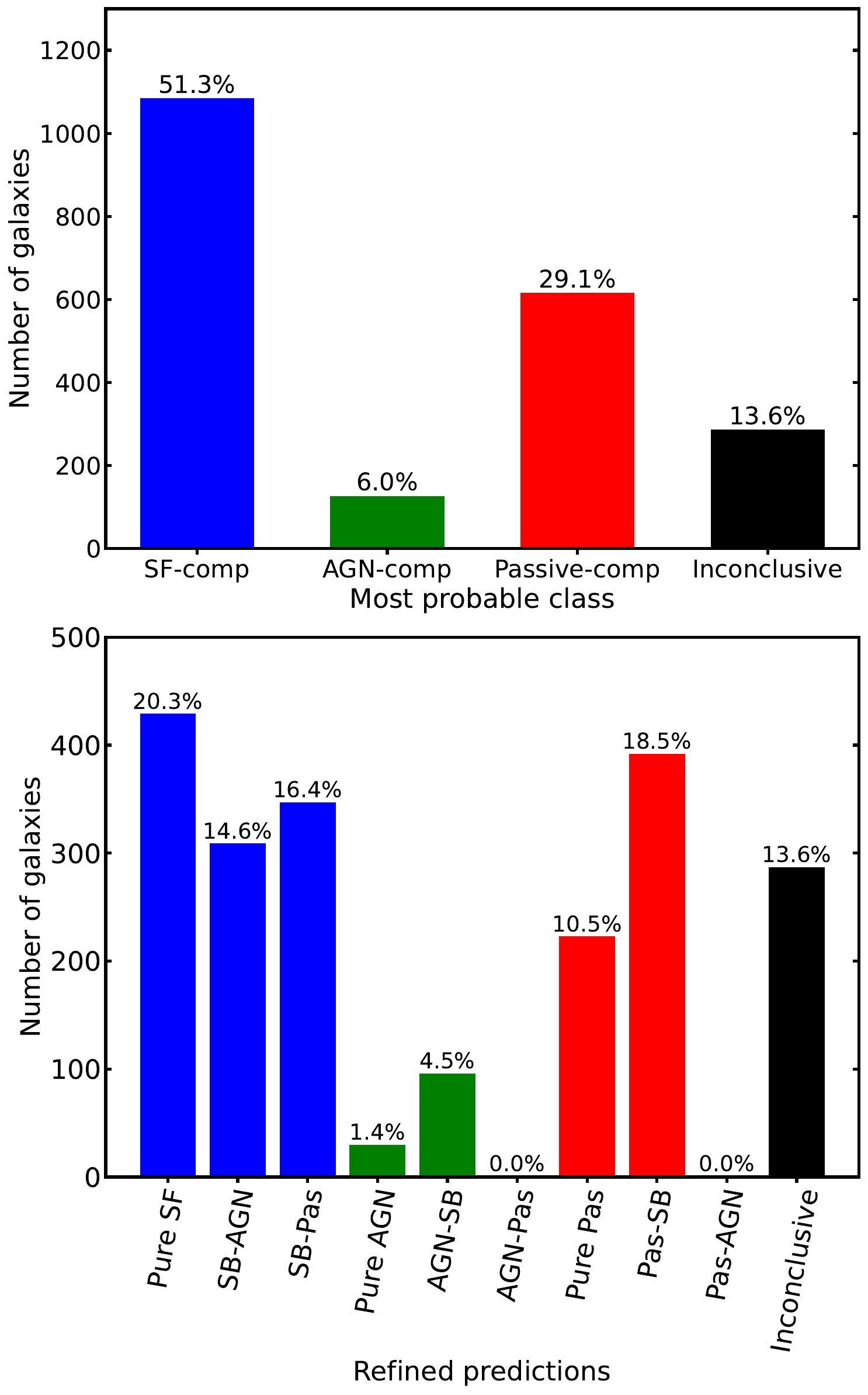}
\end{center}
\caption{The most probable (highest similarity) class among the three principal activity classes (top) and predictions based on the DONHa classification scheme (bottom) for the subsample of composite galaxies. The bars in both histograms are color-coded based on the principal class that each galaxy is predicted to resemble most. For inconclusive see Sect. \ref{dnc}. Labels: SF (star-forming), SB (starburst), AGN (active galactic nucleus), Pas (passive).}
\label{fig:prc_comp_crd}
\end{figure}

\begin{figure*}[t!]
\begin{center}
\includegraphics[scale=0.37]{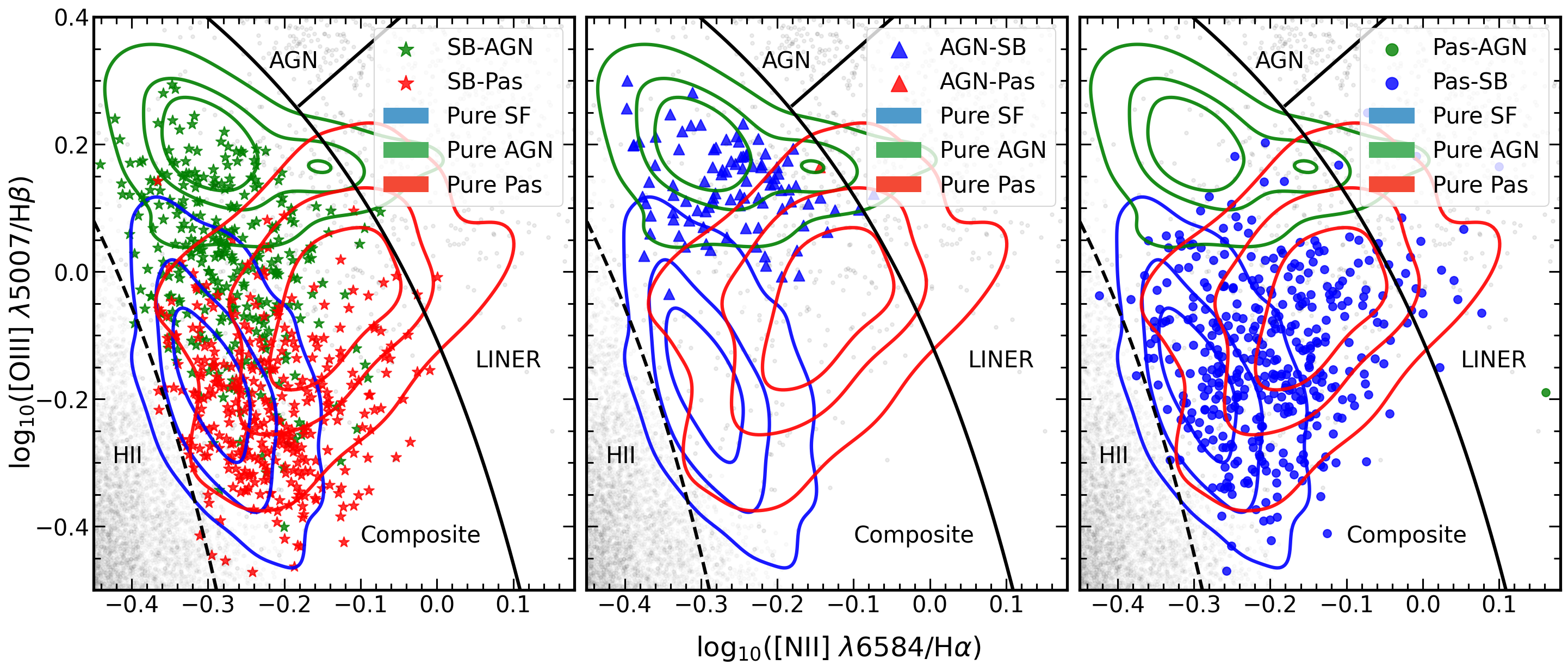}
\end{center}
\caption{[\ion{O}{III}]/H$\beta$ against [\ion{N}{II}]/H$\alpha$ flux ratio, illustrating the positions of each refined activity class as predicted by our diagnostic tool after applying it on a sample of spectroscopically selected composite galaxies. We utilized the DONHa classification scheme outlined in Table \ref{tab:tab_nwcldef}. Left, middle and right panel indicate spectra excited by star formation (SB), AGN activity and old stellar populations (Pas) respectively. Contours denote galaxies predicted to exclusively belong to one of the primary activity classes (max\_pi > 90\%), while data points represent galaxies with mixed activities (max\_pi < 90\%). In all plots symbols represent the class with the highest predicted probability while colors the class with second highest probability. Black boundary lines are the same as defined in Fig. \ref{fig:figBPT_EL}. The black dots represent the training sample for demonstrating purposes.}
\label{fig:comp_bpt}
\end{figure*}

\begin{figure}[h!]
\begin{center}
\includegraphics[width=9cm,height=13cm]{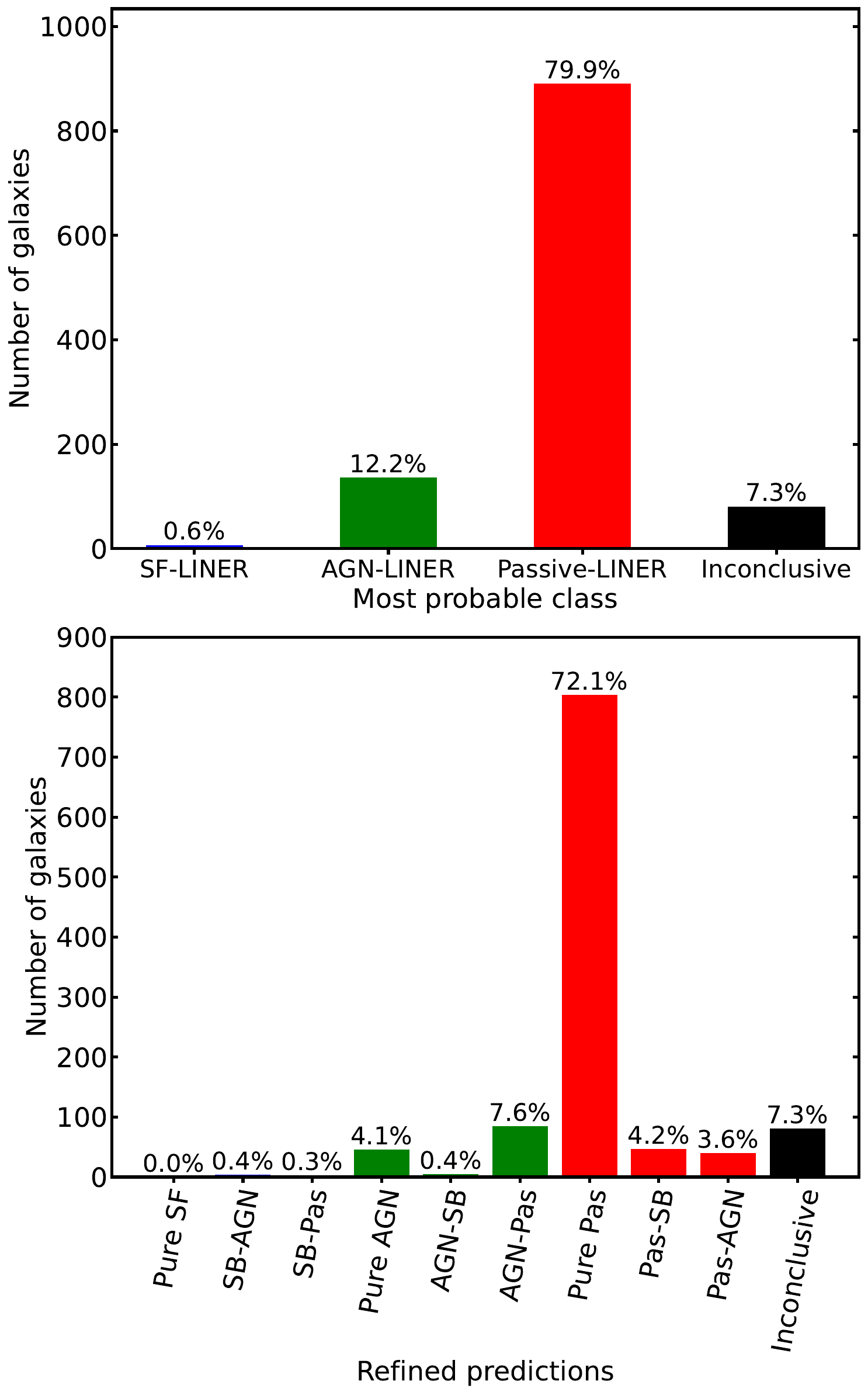}
\end{center}
\caption{The most probable (highest similarity) class among the three principal activity classes (top) and predictions based on the DONHa classification scheme (bottom) for the subsample of LINER galaxies. The bars in both histograms are color-coded based on the principal class that each galaxy is predicted to resemble most. For inconclusive see Sect. \ref{dnc}. Labels: SF (Star-forming), SB (starburst), AGN (active galactic nucleus), Pas (passive).}
\label{fig:lnr_crd}
\end{figure}

\begin{figure*}[h]
\begin{center}
\includegraphics[scale=0.34]{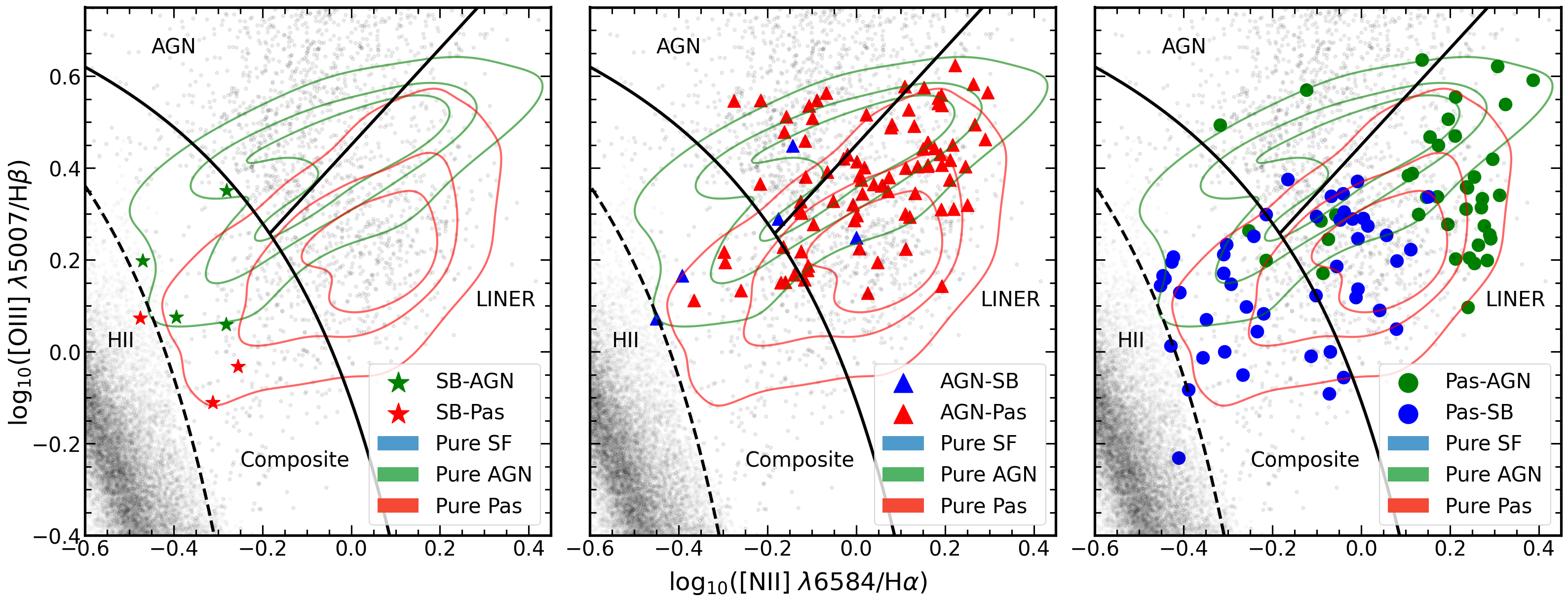}
\end{center}
\caption{[\ion{O}{III}]/H$\beta$ against [\ion{N}{II}]/H$\alpha$, illustrating the positions of each refined activity class as predicted by our diagnostic tool (DONHa) after applying it on a sample of spectroscopically selected LINER galaxies. We utilized the DONHa classification scheme outlined in Table \ref{tab:tab_nwcldef}. Left, middle, and right panels indicate spectra excited by star-formation (SB), AGN activity, and old stellar populations (Pas) respectively. Contours, symbols, and color-coding similar to Fig. \ref{fig:comp_bpt}. Black boundary lines are the same as defined in Fig. \ref{fig:figBPT_EL}. The black dots represent the training sample for demonstrating purposes.}
\label{fig:lnr_bpt}
\end{figure*}

\begin{figure*}[h]
\begin{center}
\includegraphics[scale=0.65]{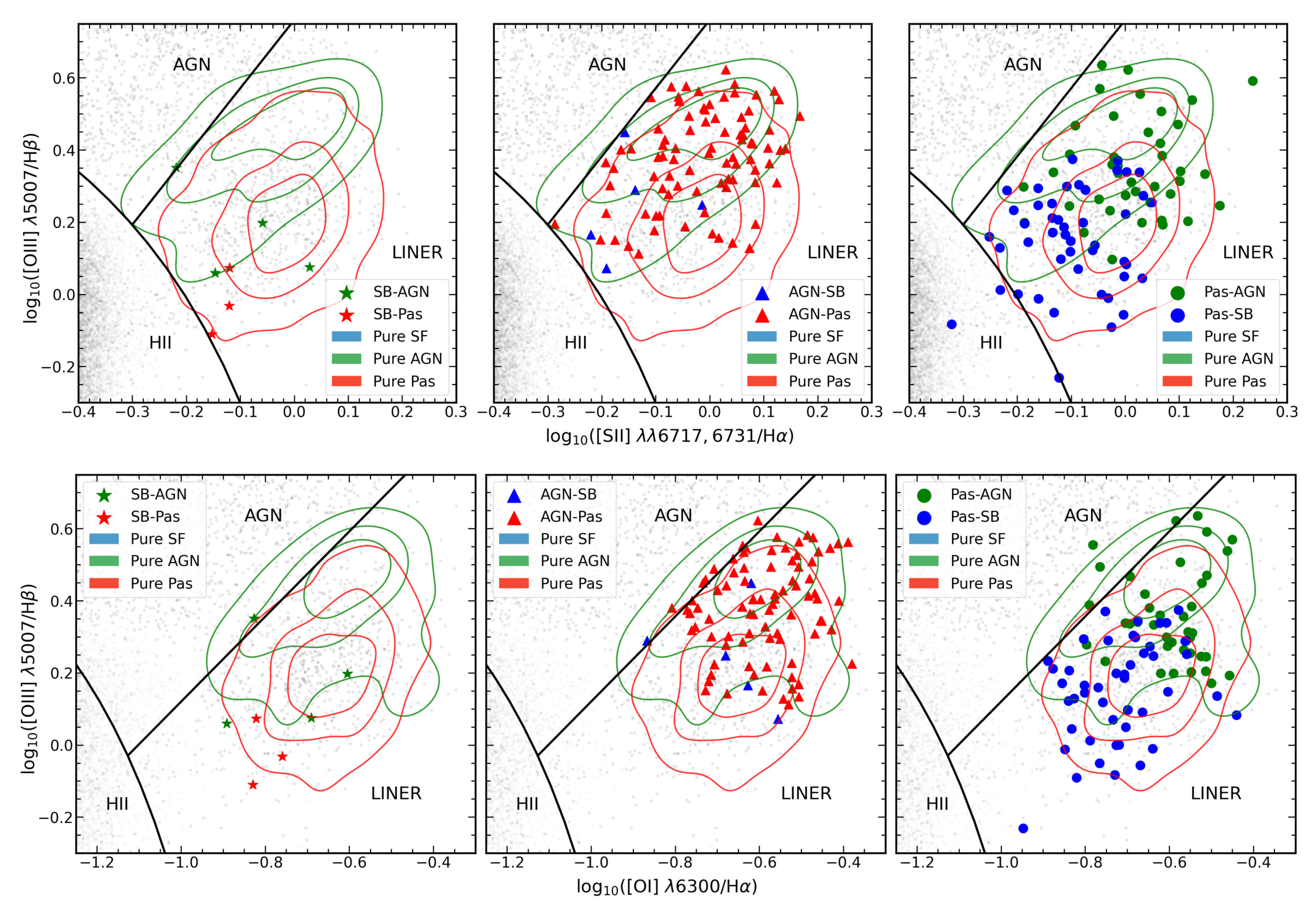}
\end{center}
\caption{Log$_{10}$([\ion{O}{III}]/H$\beta$) against log$_{10}$([\ion{S}{II}]/H$\alpha$) (top row) and log$_{10}$([\ion{O}{III}]/H$\beta$) against log$_{10}$([\ion{O}{I}]/H$\alpha$) (bottom row) are presented for the spectroscopically selected subsample of LINER galaxies. The left, middle, and right panels highlight LINER-like spectra that are found to be dominated by star-formation (SB), AGN activity and old stellar populations (Pas) respectively. Contours, symbols, and color-coding similar to Fig. \ref{fig:comp_bpt}. In all plots, black boundary lines are the same as defined in Fig. \ref{fig:figBPT_EL}. The black dots denote the training sample for demonstration purposes.}
\label{fig:lnr_bpt_SII_OI}
\end{figure*}

\begin{figure}[h]
\begin{center}
\includegraphics[width=9cm,height=7.5cm]{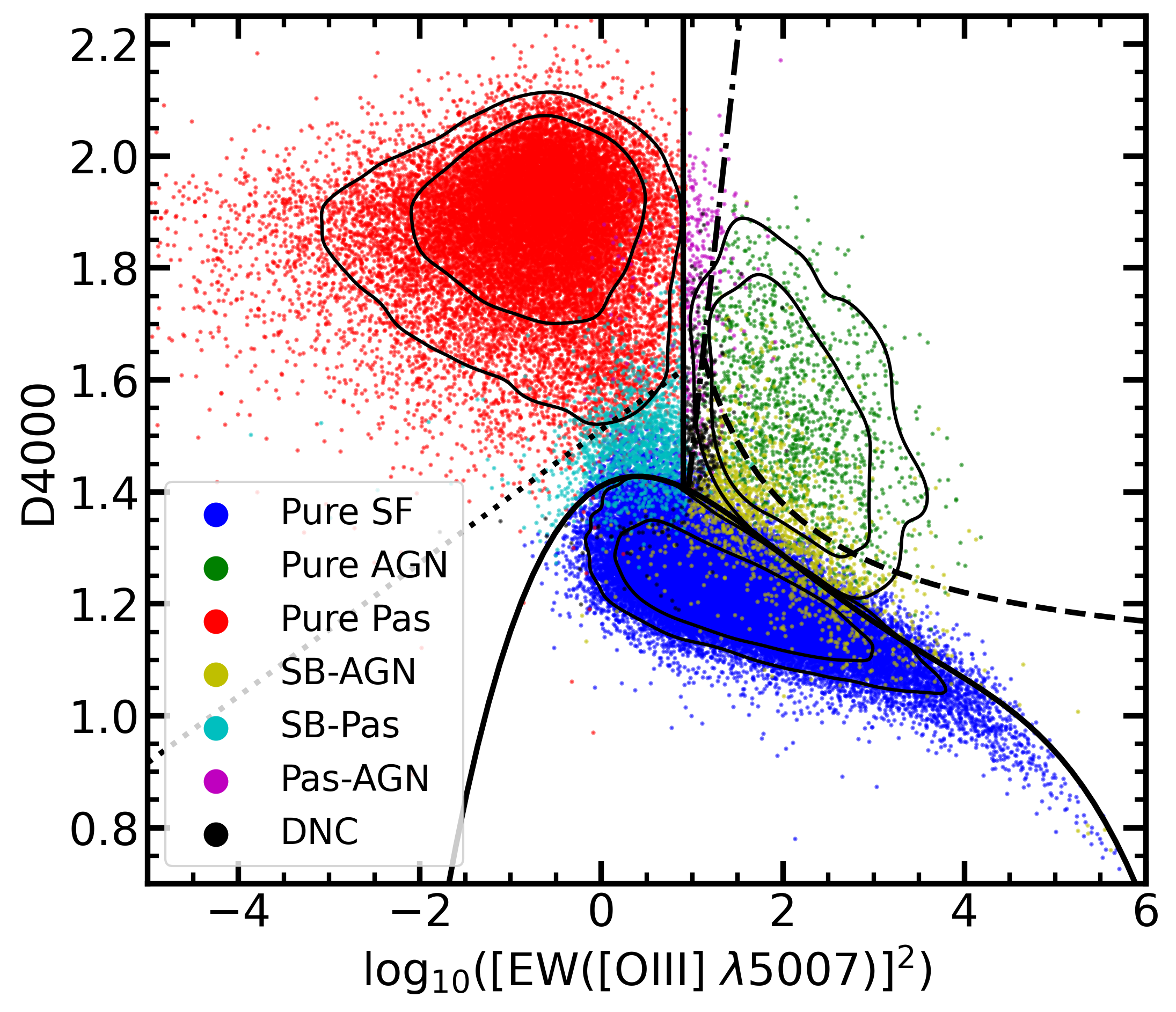}
\end{center}
\caption{The two-dimensional diagnostic diagram displays the D4000 continuum break against the log$_{10}$(EW([\ion{O}{III}])$^{2}$), the DO3 diagnostic diagram. Within this plot, we observe the spatial distribution of galaxies predominantly influenced by a single principal excitation mechanism (pure activity classes), as well as the positions of those with mixed activity classes. The black lines are the boundaries separating the different classes as described in Sect. \ref{sec45}. For the equation of each boundary see Appendix \ref{appendix2}. The contours representing each pure activity class are computed using a kernel density function and illustrate the 68\% (inner contour) and 90\% (outer contour) population density levels. Labels: pure SB (pure starburst), pure AGN (pure active galactic nucleus), pure Pas (pure passive).}
\label{fig:neo_BPT}
\end{figure}

\begin{figure}[h]
\begin{center}
\includegraphics[scale=0.4]{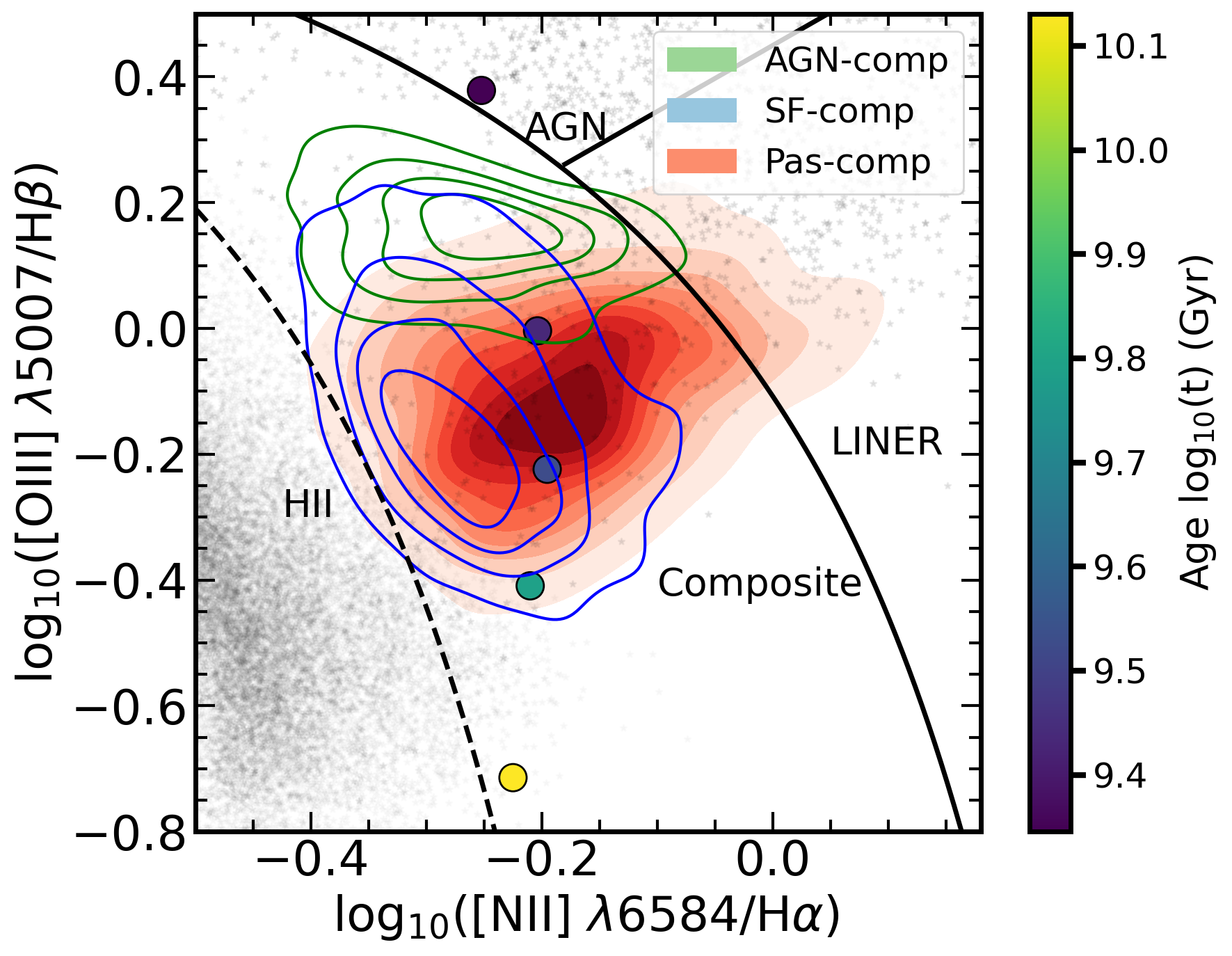}
\end{center}
\caption{Log$_{10}$([\ion{O}{III}]/H$\beta$) against log$_{10}$([\ion{N}{II}]/H$\alpha$). Within this plot, we can see the distribution of the spectroscopically selected subsample of composite galaxies that have been classified by the new diagnostic as passive-composite galaxies, i.e., composites where the ionization source originates from hot-evolved stars. The circular data points are derived from the photoionization models form the work of \cite{2019AJ....158....2B}, and they are overlaid to indicate the locations of galaxies with aging stellar populations spanning from 2 to 14 Gyr. The circles are color-coded to denote the age of the stellar populations. Black boundary lines are the same as defined in Fig. \ref{fig:figBPT_EL}. The black dots in the top plot represent the training sample of the principal classes, included for illustrative purposes.}
\label{fig:cp_Byler_bpt}
\end{figure}

\begin{figure}[h]
\begin{center}
\includegraphics[width=8.8cm,height=7cm]{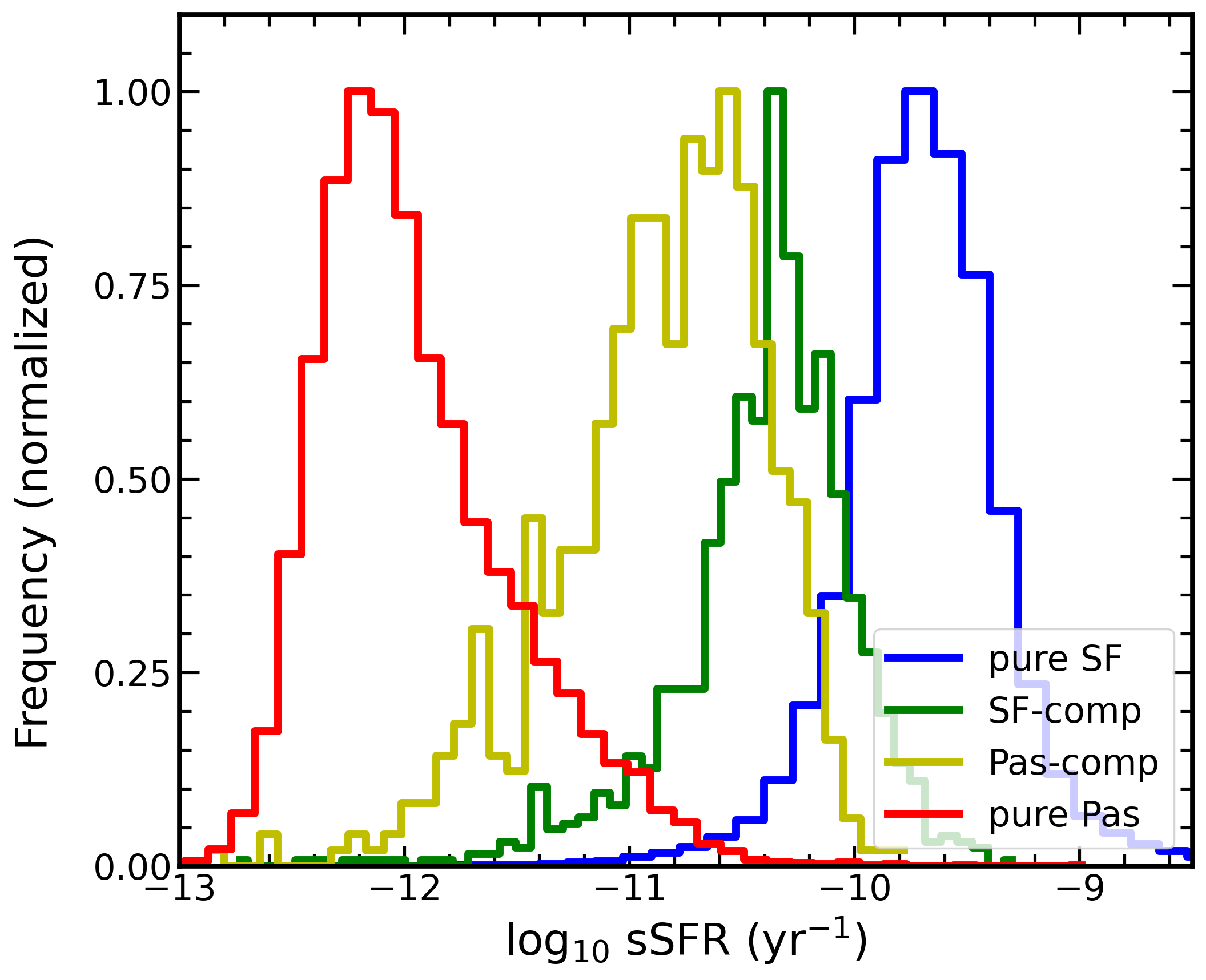}
\end{center}
\caption{The distribution of the specific star-formation rate (sSFR) is examined for composite galaxies predicted to share similar characteristics with star-forming galaxies (green) and passive galaxies (yellow). Additionally, we present the distributions of star-forming (blue) and passive (red) galaxies from our sample for illustration purposes.}
\label{fig:comp_ssfr}
\end{figure}
\noindent Indeed, we find that in our sample of passive galaxies selected this way, the contamination by spectroscopically classified star-forming or AGN galaxies is negligible ($\sim$ 0.5\% and $\sim$ 0.7\% respectively). Therefore, we opted not to remove any emission-line objects that have been spectroscopically classified as star-forming or AGN. Furthermore, we observed a significant number of optically selected LINER and composite galaxies that are located in the red sequence on the $(NUV-r)-M_{r}$ CMD. Since LINERs and composites may be powered by old stellar populations we adopt the passive classification. We prefer the $NUV-r$ definition for the red sequence over the definition based on the $(u-r)-M_{r}$ CMD since we found that the latter has significant contamination by AGN and obscured star-forming galaxies. This approach provides a continuous mapping of the different observed features, ranging from weak emission to absorption, as a result of the diverse evolutionary stages of underlying stellar populations found within the host galaxy.

Even though the gas ionization mechanism in LINER and composite galaxies is generally considered to be an active nucleus, there is growing evidence that their emission can also be attributed to hot evolved stellar populations \citep[post-AGB stars; e.g.,][]{2008MNRAS.391L..29S,2013A&A...558A..43S}. In the case of composite galaxies, the prevailing notion is that their activity originates from both an active nucleus and a star-formation component. However, as discussed in \cite{2019AJ....158....2B}, it is also possible that emission lines of these galaxies may be excited by weak residual star formation aided by ionization from hot-evolved stellar populations. Even though these two sub-populations of LINER and composite galaxies exhibit emission lines that, under different circumstances, would characterize them as active galaxies, the primary ionization mechanism derives from old stellar populations. By incorporating objects with faint emission lines into the training sample, we do not bias our diagnostic against galaxies with excitation by old stellar populations.

% \begin{figure}[h!]
% \begin{center}
% \includegraphics[scale=0.39]{plots/g_r_CMD_contours_pas.png}
% \end{center}
% \caption{CMD diagram of $g-r$ against $M_{r}$. The red dots represent the selected sample of passive galaxies. The grey dots depict the star-forming galaxies in our sample for scale. We see that the galaxies we have selected as passive form a distribution that matches the locus of red sequence of galaxies \citep{2004MNRAS.351.1151B}. The contours illustrate the density of the objects.}
% \label{fig:figCMD}
% \end{figure}

% Because as of now, no single diagnostic can offer a unified classification scheme, in this section and in Sect. \ref{sec:22}, we adopted two distinct approaches for selecting passive and active galaxies, recognizing the fundamental differences in their nature. For passive galaxies, the selection process involved utilizing the color-magnitude diagram (CMD) of $NUV-r$ color against $r$-absolute magnitude ($M_{r}$). To ensure the robustness of the classification, we implemented a signal-to-noise (S/N) selection criterion. Specifically, any passive galaxy with S/N > 3 in the $NUV-r$ color was included in the final sample.

Finally, we applied extinction and k-corrections to all utilized photometry. The optical $u$ and $r$ SDSS colors of the galaxies are also corrected for galactic dust extinction, following the \cite{1989IAUS..135P...5C} extinction law with $R_{V} = 3.1$, and $E(B-V)$ values obtained from the dust maps of \cite{1998ApJ...500..525S}. The NUV band was corrected for reddening effects using the extinction coefficients provided by \cite{2013ApJ...771...68P}. Given the substantial range of distances covered by galaxies in our sample (up to $z=0.08$), we also performed k-corrections in the NUV and r bands using the k-correction {calculator}\footnote[1]{\url{http://kcor.sai.msu.ru/}} based on the methods presented in \cite{2010MNRAS.405.1409C} and \cite{2012MNRAS.419.1727C}.

\subsection{Data processing and final sample} \label{sec24}

Our focus is on training a diagnostic for the three principal mechanisms of gas excitation. For this reason from the emission-line classification we only consider these three principal classes: star formation, active nucleus, and old stellar populations. All other active galaxy classes, such as composite and LINER galaxies, are intentionally excluded from the training dataset. The composition breakdown per class within the final sample, intended for algorithm training, is detailed in Table \ref{table:tab_fs}.

In Fig. \ref{fig:figBPT_EL}, we depict the projection of our sample on the traditional two-dimensional BPT diagrams illustrating the distribution of the training set for each of the three classes: SF, AGN, and passive. The projections are shown with respect to log$_{10}$([\ion{O}{III}] $\lambda5007$/H$\beta$) against log$_{10}$([\ion{N}{II}] $\lambda6584$/H$\alpha$), log$_{10}$([\ion{S}{II}] $\lambda \lambda6717,6731$/H$\alpha$), and log$_{10}$([\ion{O}{I}] $\lambda6300$/H$\alpha$).

These diagrams also include the subset of passive galaxies with S/N > 3 in all emission lines used for the plots. As anticipated, a non-negligible fraction of the passive galaxy sample is absent from these diagrams due to their minimal or nonexistent emission lines. However, the presence of passive galaxies in these diagrams ensures that our passive galaxy sample is not biased against early-type galaxies hosting an AGN, residual star formation, or photoionized regions by post-AGN stars. As expected, these are found in the locus of composite and LINER galaxies \citep[c.f., ][]{2010MNRAS.403.1036C}.

\subsection{Feature selection}

Activity diagnostics in the optical band are traditionally based on emission-line ratios to characterize galactic activity. However, this introduces a bias, as early-type galaxies tend to possess limited or negligible reserves of dust and gas, thereby hindering their ability to generate such lines and as a result excluding them from these diagnostics. To address these limitations, we propose the utilization of equivalent width measurements of a spectral line instead of its flux. This approach aims to address these issues and establish a self-consistent diagnostic method applicable seamlessly to galaxies exhibiting emission on absorption lines.

Our next step is to determine the minimum number of spectral lines required to identify the dominant ionization mechanism within the host galaxy. In our pursuit of identifying these optimal features, we commence by contemplating both astrophysical and practical considerations for each potential feature selection. Beginning from an astrophysical standpoint, our motivation is partially rooted in the physics underlying emission-line diagnostic methods. We specifically consider the EWs of the spectral lines present in BPT diagrams: H$\alpha$, [\ion{N}{II}] $\lambda$6584, [\ion{O}{III}] $\lambda$5007, [\ion{S}{II}] $\lambda \lambda$6717,6731, and [\ion{O}{I}] $\lambda$6300. There is a strong physical reasoning for this choice, as we anticipate that the EWs of singly or doubly ionized oxygen will be higher in an AGN environment compared to an \ion{H}{II} region \citep{1981PASP...93....5B}. This arises from the fact that the UV radiation in an AGN environment tends to be more energetic than that typically encountered in an \ion{H}{II} region.

From a practical perspective, we opted for strong spectral features that are generally straightforward to observe and measure, such as the Balmer lines and strong forbidden emission lines (e.g., [\ion{O}{III}] $\lambda$5007). In Fig. \ref{fig:feature_dist} we see the distributions of all the features we considered as potential discriminating features for the development of our diagnostic. However, due to the correlation between the H$\alpha$ and H$\beta$ lines including both is redundant. Consequently, we choose to utilize H$\alpha$ because of its larger EW, making it applicable to a broader range. Similarly, we observe that EW([\ion{S}{II}] $\lambda \lambda$6717,6731) and EW([\ion{O}{I}] $\lambda$6300) exhibit very similar behavior to EW(H$\alpha$) and the EW([\ion{N}{II}] $\lambda$6584). This observation suggests that the inclusion of the EW of these lines, [\ion{S}{II}] $\lambda \lambda$6717,6731 and [\ion{O}{I}] $\lambda$6300, in our diagnostic scheme may not provide significant additional information for discriminating the primary activity classes. Therefore we adopt the EW of H$\alpha$, [\ion{O}{III}] $\lambda$5007, and [\ion{N}{II}] $\lambda$6584 spectral lines.

The features we discussed in the previous paragraph are good indicators of the activity of a galaxy. However, these activity indicators alone are not capable of identifying whether a galaxy hosts primarily old or young stellar populations which can produce ionizing continua with hardness between that of a young stellar population and an AGN \citep{2010MNRAS.403.1036C}. So, in addition to these features, we also require a feature that conveys information regarding the age of the stellar populations. Of all stellar absorption line indicators, the D4000 is the only one that shows the weakest metallicity dependence \citep{1997A&A...325.1025P}. The amplitude of the D4000 index is primarily influenced by massive stars in the main sequence (MS), and therefore, it exhibits nearly monotonic increase with the age of the stellar populations. The definition of D4000 we have adopted is from the work of \cite{1999ApJ...527...54B}, which uses narrower bands than the one typically defined by \cite{1983ApJ...273..105B}. This enables us to include a larger number of galaxies while simultaneously being less sensitive to reddening effects.

Concluding, we find that by using the EW of H$\alpha$, [\ion{O}{III}] $\lambda$5007, and [\ion{N}{II}] $\lambda$6584 lines along with the D4000 continuum break index, we can define a diagnostic tool that fulfills all proposed criteria. The distributions of each feature per class are presented in Fig. \ref{fig:feature_dist}.

\section{The classifier} \label{sec3}
\subsection{The random forest} \label{rf_ds}

The complexity of this problem necessitates the utilization of a versatile algorithm capable of effectively discriminating between galaxy activity classes. We have determined that the random forest algorithm \citep{2014arXiv1407.7502L}, is best suited to address the requirements of this classification task. Our choice of this specific algorithm is motivated by its inherent flexibility, as it allows the adjustment and fine-tuning of numerous parameters to align with the specific demands of the problem at hand. Furthermore, the random forest algorithm is well-known for producing robust results, as its training process is generally unaffected by outliers. Another advantage lies in its straightforward and intuitive operation, which simplifies the implementation and interpretation of results. Also, along with the classification labels the random forest calculates the probability of the classified object to belong in each one of the considered classes.

\subsection{Implementation}\label{impl}

We implement the random forest algorithm using the \texttt{RandomForestClassifier} from the \texttt{scikit-learn} Python 3 package \citep{scikit-learn}, version 1.1.1. Our algorithm is provided with four features: EWs of [\ion{O}{III}] $\lambda$5007, [\ion{N}{II}] $\lambda$6584, and H$\alpha$, the D4000 continuum break, and the activity class of each object. These features enable the algorithm to be trained to classify galaxies into the three primary activity types: star-forming, AGN, and passive.

Following the selection of the optimal values of the hyperparameters (for more details, see Appendix \ref{appendix1}), the subsequent step is training the algorithm. We employ the data that were selected based on the specific criteria outlined in Sect. \ref{sec24}. The entire sample is divided into two subsets: the training set and the test set, with a 70\%-30\% split respectively. To ensure uniformity within the two subsets, we perform a stratified split ensuring that each subset contains the same fraction of each class. Before splitting the data, we randomly shuffle them to guarantee homogeneity within the two subsets. The training set, comprising the majority (70\%) of the data, is utilized for training the algorithm, serving as the data from which the decision trees are constructed. On the other hand, the test subset, which accounts for the remaining 30\% of the sample, is solely used for evaluating the performance of the algorithm. This approach ensures that the algorithm generalizes well and performs effectively on unseen data, extending its utility to datasets beyond the one used for training.

\subsection{Performance metrics} \label{p33}

To visualize the performance, we calculate a confusion matrix based on a subset of objects for which we already have their ground-truth classifications (the test set). A perfect classifier is characterized by a confusion matrix with elements only along its primary diagonal, indicating that every prediction matches the true class. However, if the classifier has misclassified instances, the off-diagonal elements will be non-zero, offering a detailed view of the misclassified objects.

In addition to the confusion matrix, we use several other performance metrics as described in Table 2 in \cite{2023A&A...679A..76D}. In particular, we make use of the accuracy, the recall, and the F$_{1}$ score.

\section{Results} \label{res}

\subsection{Performance on the principle activity classes}
\label{prf}

We evaluate the performance of our model by calculating its accuracy by splitting the whole sample into ten subsamples. Each time we use nine out of the ten subsamples (folds) to train it and one to evaluate the accuracy. We repeat the training process by replacing each time the test set with one of the training sets. This process repeats until all the folds have been in the position of the test fold once. This is known as K-fold cross-validation. We find that the overall accuracy we achieve is 0.989$\pm$0.004. The low standard deviation suggests a robust classifier as its accuracy score is independent of the training data.

\begin{table}[h]
\caption{Report of performance scores calculated on the test sample for each galaxy class using three different metrics.}
\centering
\begin{tabular}{l c c c c }
\hline\hline
Class & Precision & Recall & F$_{1}$-score & Galaxies \\
\hline
Star-forming & 1.00 & 1.00 & 1.00 & 10822\\
AGN & 0.84 & 0.98 & 0.91 & 384\\
Passive & 1.00 & 0.99 & 0.99 & 6399 \\
\hline
\end{tabular}
\label{tab:tabpsc}
\end{table}

Table \ref{tab:tabpsc} provides an overview of the performance scores for each class. The scores for all classes are nearly perfect. The high recall scores for each class indicate that the classifier can correctly retrieve nearly all objects of each class, showcasing its high level of completeness. Furthermore, the high precision scores imply that there are only a few instances where objects have been incorrectly predicted to belong to a different class than their true class. This demonstrates that the contamination within each class is minimal. These results collectively indicate the effectiveness of the diagnostic in accurately identifying the galaxies that are representative of the three principal gas excitation mechanisms (i.e., young stars, active nucleus, and old stellar populations).

In Fig. \ref{fig:figurercn}, we present the confusion matrix for this diagnostic, which was calculated using the objects from the test subset. It is clear that the confusion matrix exhibits a nearly diagonal pattern. Misclassification instances for star-forming and passive galaxies are negligible. There, only a small fraction of misclassified AGN galaxies (1.6\%) as star-forming.

\begin{table*}[h]
\caption{Definitions and selection criteria for all activity classes based on the refined classification scheme (DONHa diagnostic).}
\centering
\begin{tabular}{l l l}
\hline\hline
Class name & Definition  & Criterion\\ 
\hline
Pure starburst & Dominant photoionization source & max\_$p_{i}$ $\geq$ 90\% to be starburst \\& young massive stars.\\ \\
Pure AGN & Dominant photoionization source & max\_p$_{i}$ $\geq$ 90\% to be AGN \\& active nucleus.  \\ \\
Pure passive & Dominant photoionization source & max\_$p_{i}$ $\geq$ 90\% to be passive \\& hot evolved stellar populations. \\ \\
Starburst-AGN & Star-formation and an active nucleus. & max\_$p_{i}$ < 90\% to be starburst, while the second \\& Dominant of the two is star-formation. & higher predicted probability is AGN. \\&  \\
AGN-starburst & Star-formation and an active nucleus. & max\_$p_{i}$ < 90\% to be AGN, while the second \\& Dominant of the two is an active nucleus. & higher predicted probability is starburst. \\&  \\
Starburst-passive & Star-formation and hot-evolved stars. & max\_$p_{i}$ < 90\% to be starburst, while the second \\& Dominant of the two is star-formation. & higher predicted probability is passive. \\&  \\
Passive-starburst & Star-formation and hot-evolved stars. & max\_$p_{i}$ < 90\% to be passive, while the second \\& Dominant of the two is hot-evolved stars. & higher predicted probability is starburst. \\&   \\
AGN-passive & Hot-evolved stars and an active nucleus. & max\_$p_{i}$ < 90\% to be AGN, while the second \\& Dominant of the two is active nucleus. & higher predicted probability is passive. \\&   \\
Passive-AGN & Hot-evolved stars and active nucleus. & max\_$p_{i}$ < 90\% to be passive, while the second \\& Dominant of the two is hot-evolved stars. & higher predicted probability is AGN. \\&   \\
Inconclusive & All three classes have similar probabilities & See Sect. \ref{dnc} for a discussion. \\
\hline
\end{tabular}
\label{tab:tab_nwcldef}
\tablefoot{The max\_$p_{i}$ refers to the probability of the highest-ranking class assigned to a galaxy by our classifier. In the case of mixed activity classes, the first class in each label represents the primary class (with the highest predicted probability), while the second class represents the secondary class (with the second-highest predicted probability).}
\end{table*}

\subsection{Feature importance}

While all initially selected features may appear relevant, some may have a greater impact than others. Identifying the most important features allows us to identify any redundant features that should be eliminated. This reduction in feature complexity not only results in a more efficient classifier but also enhances its adaptability to a broader number of datasets.

Fig. \ref{fig:ftip} provides a feature importance plot, extracted during the training process of the algorithm. Based on the feature importance we find that the EW of H$\alpha$ is the most crucial feature. The second most important feature, as ranked by importance, is the D4000 index. Notably, all features exhibit similar importance, indicating the robustness of the feature scheme. A closer examination of this plot reveals that the EW of [\ion{N}{II}] is ranked last and exhibits a high standard deviation. This hints at the possibility of a feature with low significance, suggesting redundancy. However, the removal of the EW of [\ion{N}{II}] from the feature scheme resulted in a significant decrease in performance, particularly for the AGN galaxies.

\subsection{Mixed-activity classes and predicted probabilities}
\label{c43}

So far, we have developed a diagnostic that focuses on three classes representing the principal mechanisms of gas excitation in galaxies. However, often different excitation mechanisms can co-exist (e.g., star-formation, AGN, and old stellar populations). One such example is composite galaxies, where both an active nucleus and ongoing star-formation processes may be present simultaneously. Another scenario for composite galaxies involves gas excitation by hot-evolved stars alongside an active star-formation component. Additionally, the class of LINER galaxies represents another group with mixed activity, which may be powered by an AGN (even displaying broad emission lines; \cite{1997ApJS..112..315H} or old stellar populations \citep{1994A&A...292...13B,2008MNRAS.391L..29S,2013A&A...555L...1P}.

Our analysis so far has resulted in a tool that is able to clearly separate the galaxies that are dominated by one of the principal activity classes. This is supported by the well-defined distributions in the four-dimensional feature space, showcasing excellent separation among them. However, it is important to note that there is some overlap among these distributions. This overlap is a critical aspect as it allows us to map the entire range of distributions of the different classes in the four-dimensional feature space and identify outliers within each of the three classes. It is expected that the transition and mixing from one pure principal activity class to another is a continuous process within our four-dimensional feature space. As a result, the diagnostic possesses the capability to deconstruct mixed activity classes into their principal activity components based on the resemblance of the observed spectral features of their distributions for the different excitation mechanisms, as they are mapped in the four-dimensional feature space we consider. This decomposition ability is a direct consequence of the random forest algorithm as well as the chosen feature space, as mixed classes naturally exist in the intermediate region between the principal activity classes.

% The decomposition of galaxy activity is accomplished by utilizing the predicted probabilities generated by the random forest classifier. When the diagnostic tool is applied to a mixed-activity galaxy, each decision tree within 

The random forest functions as a probabilistic classifier, employing a 'voting' process where each tree in the ensemble independently decides the class of each object. Subsequently, an object is assigned to the class that most closely resembles it, along with the associated probabilities indicating its likelihood of belonging to each of the considered classes. As a result, galaxies that are dominated by a single activity mechanism will have a high first-ranking predicted probability of belonging to the assigned activity class. However, in the case of mixed-activity galaxies, where their activity cannot be uniquely attributed to a single mechanism, all predicted probabilities for each class will tend to be similar. In other words, due to the nature of mixed-activity galaxies, they are likely to share characteristics with multiple principal classes, leading to a lower maximum predicted probability for their first-ranked class.

\subsection{Definition of a refined classification scheme} \label{sec44}

We can leverage the predicted probabilities generated by the random forest to expand our analysis to the mixed activity classes. For example an object that has a non negligible probability to be classified into more than one classes can be considered as a mixed class object since it is located in the overlapping region between the locus of pure classes in the four-dimensional parameter space we consider. This way we can establish selection criteria for each activity class through an analysis of the first- and second-rank predicted probabilities for each galaxy. This allows us to include in our scheme and characterize the mixed classes.

% We start by defining the pure/principal activity classes which include galaxies that their activity is dominated by one of the main gas excitation mechanisms.

This way the activity classes from this new classification scheme (hereafter DONHa diagnostic for D4000, [\ion{O}{III}], [\ion{N}{II}], and H$\alpha$) can be categorized into two main groups: pure (or principal) and mixed-activity classes. Beginning with the pure activity classes, we retain the three classes of star-forming (or starburst), AGN, and passive, introduced in Sect. \ref{sec:22} and \ref{sec:23} which serve as representatives of the three primary gas excitation mechanisms: star formation, active nucleus, and old-stellar populations, respectively. In our analysis so far, we were interested in the predicted probabilities of each galaxy rather than just the classification output from the diagnostic. We use these in order to define new selection thresholds for each of primary activity class. 

To adjust the probability selection thresholds, we analyzed the predicted probabilities for each of the three principal classes individually, utilizing the test sample (Sect. \ref{impl}). As these galaxies were rigorously selected, we assume they are dominated by only one activity mechanism (i.e., pure classes) and represent pure activity classes. Subsequently we examined the probability of the first ranked class of the population of each of the three pure classes (i.e., star-forming, AGN, and passive galaxies). We find that the lowest 90\% percentile of the probability distribution for all the first ranking classes is higher than 90\%. This means that more than 90\% of the objects belonging in each of the classes we consider have a classification probability higher than 90\%. We denote this probability as max\_$p_{i}$ since it is the highest probability (corresponding to the highest ranking class) for each object.

The second group of activity classes describes the mixed-activity galaxies, characterized by a lower first-rank probability of resembling a pure class. Therefore, if the max\_$p_{i}$ falls below 90\%, we classify the galaxy as a mixed-activity one. For such objects, we consider not only the first- but also the second-rank predicted probability. This way we characterize each mixed-activity galaxy as having properties of two principal activity classes that describe the two dominant gas excitation mechanisms present in each mixed activity class. This approach is akin to how clustering algorithms such as K-means or Gaussian Mixture Models (GMM) operate. These algorithms categorize objects into groups based on their relative proximity within the feature space (two examples include \cite{1998ApJ...508..314M} and \cite{2019MNRAS.485.1085S}). By considering the first and second highest predicted probabilities, we can effectively characterize objects that fall between the locus of the considered classes (e.g., composite) based on their similarity to the main classes originally considered.

The names we assign to the additional activity classes are derived by pairing pure classes, resulting in six new activity classes through permutations. In addition, we place the principal class with the highest predicted probability in the first position of the label, and the class with the second highest probability in the second position. For instance, in this refined classification scheme, the mixed activity class "starburst-AGN" differs from "AGN-starburst". Although both labels convey that these two galaxy classes are primarily characterized by star formation and AGN processes, "starburst-AGN" describes a galaxy in which the dominant source of excitation originates from star formation, while "AGN-starburst" characterizes a galaxy in which the dominant source of excitation results from an active nucleus. Comprehensive definitions and selection criteria for the classes in the refined (DONHa) activity classification scheme are provided in Table \ref{tab:tab_nwcldef}. It is essential to note that the dominant source of ionization is determined based on the similarity of an object to AGN, SF, and passive galaxies in the four-dimensional space considered here, rather than the predominant flux of ionizing photons determined through SED analysis.

\subsection{Decomposing LINER and composite galaxies} \label{sec44}

Building upon the analysis and performance evaluation of our diagnostic described in the two previous sections, we proceed to utilize it for decomposing the classes of composite and LINER galaxies into their principal gas excitation components. We follow two approaches in classifying these galaxies and subsequently evaluate the results.

First, we apply the diagnostic tool to a sample of composite and LINER galaxies, drawn from our parent sample described in Sect. \ref{datasample}. They are selected by applying the SoDDA diagnostic of \cite{2019MNRAS.485.1085S}. We then apply our diagnostic defined in Sect. \ref{sec3} to classify these objects into one of the three principal classes, allowing us to characterize them based on their similarity to one of these classes. For instance, a galaxy that closely resembles a star-forming galaxy (i.e., with EWs of the diagnostic lines and D4000 are closer to the locus of the SF galaxies) will be classified as an SF-composite. The rest of the mixed activity galaxies are characterized as AGN-composite, passive-composite, SF-LINER, AGN-LINER, and passive-LINER. As discussed in the previous section the first component of the name indicates the principal activity class to which the galaxy bears the greatest resemblance, while the second component indicates its spectroscopic classification from SoDDA.

\begin{table}[h]
\caption{Class predictions following the implementation of the new diagnostic on the composite galaxy sample.}
\centering
\begin{tabular}{l c c }
\hline\hline
RF predicted class & Percentage (\%) & Galaxies  \\
\hline
SF-composite & 51.3 & 1085 \\
AGN-composite & 6.0 & 127 \\
Passive-composite & 29.1 & 616 \\
Inconclusive & 13.6 & 287 \\
\hline
Total & 100.0 & 2115 \\
\hline
\end{tabular}
\label{tab:tabcppr}
\tablefoot{The first column denotes the predicted class, determined by the random forest based on the most probable class. The second and third columns indicate the percentage and count of objects relative to the total population of spectroscopically classified composites, respectively. For inconclusive see Sect. \ref{dnc}.}
\end{table}

Based on this analysis, for the composite galaxies, it is found that $\sim51\%$ of them are predicted as being dominated by star-formation, $\sim6\%$ as being dominated by an active nucleus, and $\sim29\%$ as being dominated by hot evolved stars. For the case of LINER galaxies, we find that about $\sim12\%$ are predicted as being dominated by AGN activity and $\sim80\%$ by hot-evolved stars. LINER galaxies predicted as star-forming dominated are almost nonexistent. In Tables \ref{tab:tabcppr} and \ref{tab:tablnrpcr} we summarize these results. A small fraction of these objects (13.3\% of composite and 7.3\% of LINER galaxies) had almost equal probabilities to belong to all three principal classes. We characterize the result of their classification as inconclusive (see Sect. \ref{dnc} for a detailed discussion).

\begin{table}[h]
\caption{Class predictions following the implementation of the new diagnostic on the LINER galaxy sample.}
\centering
\begin{tabular}{l c c}
\hline\hline
RF predicted class & Percentage (\%) & Galaxies  \\
\hline
SF-LINER & 0.6 & 7 \\
AGN-LINER & 12.2 & 136 \\
Passive-LINER & 79.9 & 891 \\
Inconclusive & 7.3 & 81 \\
\hline
Total & 100.0 & 1115 \\
\hline
\end{tabular}
\label{tab:tablnrpcr}
\tablefoot{The first column denotes the predicted class as determined by the random forest. The second and third columns indicate the percentage and count of objects relative to the total population of LINERs, respectively. For inconclusive see Sect. \ref{dnc}.}
\end{table}

We can further analyze these results by exploring the locus of these different subclasses on the standard emission-line ratio diagnostic diagrams. In the ([\ion{O}{III}] $\lambda$5007/H$\beta$ against [\ion{N}{II}] $\lambda$6584/H$\alpha$) plot, the composite galaxies occupy the area between the two lines of \cite{2003MNRAS.346.1055K} and \cite{2001ApJ...556..121K}.

In Fig. \ref{fig:merged_decomp_liner}, it can be observed that the classes assigned by the new diagnostic tool form distinct clouds, with the center of each distribution being distinguishably different from the others. On that diagram, the location of the composite galaxies that are dominated by star-formation processes is just above the star-forming cloud and tangential to the \cite{2003MNRAS.346.1055K} line. Correspondingly, the AGN-composite predicted galaxies are found in the upper part of the composite population, close to the theoretical line of extreme starburst defined by \cite{2001ApJ...556..121K}. For the passive-composite galaxies, we see that their distribution is wider, extending to the area of objects with strong low-ionization emission lines (LINERs). Interestingly, they follow the trend described in \cite{2019AJ....158....2B} for galaxies with contributions from an ionizing component of older hot stellar populations.

Following this observation, we next apply the DONHa classification scheme described in Table \ref{tab:tab_nwcldef} in order to further refine our classification and to identify the primary and secondary activity classes of mixed activity objects. The results of this analysis are shown in Table \ref{tab:RF_COMP_LNR_refnd} and Fig. \ref{fig:prc_comp_crd}.

The results of the classification of the sample of composite galaxies based on two classification schemes are shown in Fig. \ref{fig:prc_comp_crd}. The top histogram of Fig. \ref{fig:prc_comp_crd} shows the predictions based on the likelihood of similarity of a composite galaxy to one of the three principal activity classes. In the bottom histogram of the same figure, we show the classification based on the more refined activity classification scheme. In Fig. \ref{fig:comp_bpt}, we project these composite galaxies onto the BPT plot to observe the location of each one of the refined activity classes.

We repeat this analysis on the subsample of LINER galaxies. In Table \ref{tab:RF_COMP_LNR_refnd} we show the results obtained by the DONHa classification scheme (see Table \ref{tab:tab_nwcldef}). In Fig. \ref{fig:lnr_crd}, we present the results of the classification obtained by discriminating them based on their similarity to the three principal activity classes (same approach as for composites, top histogram of Fig. \ref{fig:lnr_crd}) and with the DONHa classification scheme that also considers the mixing of the different gas excitation mechanisms present in the host galaxy (bottom histogram of Fig. \ref{fig:lnr_crd}). In Fig. \ref{fig:lnr_bpt}, we plot the subsample of LINER galaxies on the standard BPT plot. It is known that the [\ion{S}{II}] doublet and the [\ion{O}{I}] are good probes of low-ionization sources. For this reason, in Fig. \ref{fig:lnr_bpt_SII_OI}, we plot the subsample of LINERs on [\ion{O}{III}]/H$\beta$ against [\ion{S}{II}]/H$\alpha$ (top row) and on [\ion{O}{III}]/H$\beta$ against [\ion{O}{I}]/H$\alpha$ (bottom row) plots. The classification labels we used were assigned by the DONHa activity model. In that figure we observe a distinct difference in the center of the distribution for LINERs predicted to be powered purely by an AGN compared to those where the activity is solely attributed to old stellar populations. Furthermore, the latter distribution is noticeably below the former in both [\ion{O}{III}]/H$\beta$ against [\ion{S}{II}]/H$\alpha$ and on [\ion{O}{III}]/H$\beta$ against [\ion{O}{I}]/H$\alpha$.

% \begin{figure*}[h]
% \begin{center}
% \includegraphics[scale=0.65]{plots/BPT__SIII_OI_NII_LNR_SUB_horizontal_SUB.png}
% \end{center}
% \caption{\textbf{Plots of log$_{10}$([\ion{O}{III}]/H$\beta$) against log$_{10}$([\ion{S}{II}]/H$\alpha$) (first row) and log$_{10}$([\ion{O}{III}]/H$\beta$) against log$_{10}$([\ion{O}{I}]/H$\alpha$) (bottom row) are presented for the spectroscopically selected subsample of LINER galaxies. Labels have been assigned by our diagnostic using the refined scheme. Contours denote galaxies predicted to exclusively belong to one of the primary activity classes (max\_pi > 90\%), while data points represent galaxies with mixed activities (max\_pi < 90\%). In all plots symbols represent the class with the highest predicted probability while colors the class with second highest probability. In all plots, black boundary lines are the same as defined in Fig. \ref{fig:figBPT_EL}. The black dots denote the training sample for demonstration purposes. Labels: SB (Starburst), AGN (active galactic nucleus), Pas (passive).}}
% \label{fig:lnr_bpt_SII_OI}
% \end{figure*}

\begin{table*}[h!]
\caption{Class predictions following the application of the DONHa diagnostic to the composite and LINER galaxies.}
\centering
\begin{tabular}{l c c c c}
\hline\hline
 & \multicolumn{2}{c}{Composite} & \multicolumn{2}{c}{LINER} \\
\hline
RF predicted class & Percentage (\%) & Galaxies & Percentage (\%) & Galaxies \\
\hline
Pure starburst & 20.3 & 429 & 0.0 & 0 \\
Pure AGN & 1.4 & 30 & 4.1 & 46\\
Pure Passive & 10.5 & 223 & 72.1 & 804\\
Starburst-AGN & 14.6 & 309 & 0.4 & 4\\
AGN-starburst & 4.5 & 96 & 0.4 & 5\\
Starburst-passive & 16.4 & 347 & 0.3 & 3\\
Passive-starburst & 18.5 & 392 & 4.2 & 47\\
Passive-AGN & 0.0 & 1 & 3.6 & 40\\
AGN-passive & 0.0 & 1 & 7.6 & 85\\
Inconclusive & 13.6 & 287 & 7.3 & 81\\
\hline
Total & 100.0 & 2115 & 100.0 & 1115\\
\hline
\end{tabular}
\label{tab:RF_COMP_LNR_refnd}
\tablefoot{The first column represents the predicted class which was determined by considering not only the most probable class but also the second most likely class. These activity classes are outlined in more detail in Table \ref{tab:tab_nwcldef}. The second and third columns denote the percentage and count of objects relative to the total population of spectroscopically classified composite galaxies, respectively. Fourth and fifth columns represent same analysis as in composite but for LINER galaxies. For "inconclusive" see Sect. \ref{dnc}.}
\end{table*}

\subsection{Development of a two-dimensional diagnostic diagram} \label{sec45}

While our DONHa diagnostic provides excellent performance, for simplicity and ease of use, we present a two-dimensional diagnostic which is a projection from the original four-dimensional feature space with almost no loss in performance. This is possible due to the clear separation of the three principal activity classes in the four-dimensional feature space. We find that the best separation between the activity classes occurs when they are projected on the D4000 against the log$_{10}$(EW([\ion{O}{III} $\lambda$5007])$^{2}$). Another reason we chose these two features is that H$\alpha$ is sensitive to uncertainties from starlight subtraction and it is sometimes blended with the [\ion{N}{II}]. In Fig. \ref{fig:neo_BPT}, we present our sample of galaxies on a plot of D4000 against the log$_{10}$([EW(\ion{O}{III} $\lambda$5007)]$^{2}$) (hereafter DO3). The labels for each class have been defined by the DONHa classifier (see Table \ref{tab:tab_nwcldef}). Due to the small number of objects in the mixed classes as defined in Table \ref{tab:tab_nwcldef}, and for better clarity, we merge the mixed activity classes involving the same two classes regardless of which one is the primary and secondary. 

The different boundaries that delineate the classes are determined through the application of the support vector machine algorithm \citep[SVM;][]{cortes1995support}. Specifically, we employ the SVM algorithm provided by the \texttt{scikit-learn} package in Python 3. We train the SVM on the two-dimensional feature space comprising the D4000 and the log$_{10}$([EW(\ion{O}{III} $\lambda$5007)]$^{2}$), adopting a one-vs-rest approach. In this approach, we treat the class (or classes) on one side of the boundary we want to define as one class while considering the class (or classes) on the other side of the boundary as the other class. After training the SVM, we derive the boundaries delineating the classes and fit functions to extract these boundary equations. We show the optimal boundary equations for selecting each class in the Appendix \ref{appendix2}.

We find that the majority of the boundary lines can be accurately described by simple polynomial equations. We find that more complex equations for each boundary, such as higher-order polynomials, did not provide any advantages in terms of reliability and completeness.

From the application of the random forest diagnostic described in Sect. \ref{sec3} we find a small set of objects (13.6\% of composites and 7.3\% of LINERs) that have almost equal probabilities of belonging in each of the three main classes, making their classification inconclusive (see Sect. \ref{dnc}). Given the limited number of such instances and the dispersion of these objects in the DO3, it is not possible to determine a distinct region for inconclusive objects on the DO3 diagram. However, it is worth noting that the majority of inconclusive classifications are clustered at the center of the DO3, where the three mixed-activity classes intersect. We advise potential users of the DO3 diagram to exercise caution when interpreting the classification results of any object situated close to the intersection of all activity classes.

\begin{table*}
\centering
\caption{Comparison of the activity classification between the DO3 and the spectroscopic classification (SoDDA).}
\begin{tabular}{ccccccc}
\multicolumn{1}{l}{}               & \multicolumn{6}{c}{SoDDA \citep{2019MNRAS.485.1085S}}                               \\
\multirow{8}{*}{\rotcell{DO3}} & & SFG & AGN & LINER & Composite 
 & Total  \\ \hline
 & Pure SB    & 10175 & 10 & 35 & 641 & 10861 \\
 & Pure AGN   & 50 & 319 & 55 & 68 & 492 \\
 & Pure passive & 20 & 0 & 795 & 300 & 1115 \\
 & SB-AGN      & 391 & 51 & 23 & 414 & 879 \\
 & SB-passive   & 169 & 0 & 64 & 609 & 842 \\
 & Passive-AGN  & 17 & 4 & 143 & 83 & 247 \\
 & Total        & 10822 & 384 & 1115 & 2115 &       
\end{tabular}
\label{tabSDnBPT}
\end{table*}

\begin{table*}
\centering
\caption{Comparison of the activity classification between the DO3 and the DONHa diagnostic tool defined in Sect. \ref{sec3}.}
\begin{tabular}{ccccccccc}
\multicolumn{1}{l}{}                 & \multicolumn{8}{c}{DONHa} \\
\multirow{8}{*}{\rotcell{DO3}} &                & \multicolumn{1}{c}{Pure SB} & \multicolumn{1}{c}{Pure AGN} & \multicolumn{1}{c}{Pure passive} & \multicolumn{1}{c}{SB-AGN} & \multicolumn{1}{c}{SB-passive} & \multicolumn{1}{c}{Passive-AGN} & \multicolumn{1}{c}{Total}  \\ \hline
 & Pure SB & 10420 & 1 & 7 & 265 & 100 & 0 & 10793 \\
 & Pure AGN  & 0 & 351 & 0 & 101 & 0 & 20 & 472 \\
 & Pure passive  & 0 & 0 & 7209 & 0 & 164 & 26 & 7399 \\
 & SB-AGN  & 275 & 59 & 0 & 351 & 7 & 9 & 701 \\
 & SB-passive  & 133 & 0 & 82 & 0 & 662 & 2 & 879 \\
 & Passive-AGN  & 13 & 7 & 16 & 1 & 28 & 116 & 181 \\
 & Total  & 10841 & 418 & 7314 & 718 & 961 & 173 & 
\end{tabular}
\label{tabnBPTRF4D}
\end{table*}

To assess the effectiveness of the DO3 diagnostic diagram, we compare the derived classification on the sample defined in Sect. \ref{sec24}, with the classifications obtained with the DONHa diagnostic and with SoDDA method Sect. \ref{sec:22}. Table \ref{tabSDnBPT} shows the comparison of the DO3 diagram with the SoDDA (ground-truth) classification. We see a high degree of agreement with the two-dimensional correctly classifying the majority of galaxies, particularly in the case of the pure activity classes.

In Table \ref{tabnBPTRF4D} we compare the DO3 diagram with the DONHa diagnostic tool, showing a very good agreement between the two classification methodologies. In particular, we see excellent agreement for the pure activity classes (top right quadrant) and good performance for the mixed activity classes apart from the starburst-AGN where 37\% of them are classified as pure starbursts.

In conclusion, the DO3 diagram proves to be a simple yet effective method for classifying galaxies into the three principal activity classes and mixed activity classes, yielding results comparable to the DONHa diagnostic. Three notable advantages are evident. Firstly, the inclusion of passive galaxies in this DO3 diagram broadens its applicability to accommodate the full range of galaxy activity classes. Secondly, it successfully disentangles the ambiguity between objects whose activity results from a combination of star formation and an AGN component, or excitation from hot evolved stars, a phenomenon observed in the locus of composite galaxies on the BPT diagram. Thirdly, the fact that our DO3 diagnostic diagram requires only a short observed rest-frame spectral range (3850-5050 $\AA$) makes it widely applicable to high redshift galaxies observed in wide area spectroscopic surveys or even higher redshift objects observed with the James Webb Space Telescope (JWST).

\section{Discussion} \label{disc}

In this work, we established a diagnostic tool that can characterize the activity of galaxies based on their similarity to the locus of the three fundamental excitation mechanisms in a four-dimensional plane consisting of the EW of the H$\alpha$, \ion{O}{III}, \ion{N}{II}, lines and the continuum break at 4000 $\AA$ (D4000). The fundamental activity classes are star-forming, AGN, and passive galaxies. The advantage of using equivalent widths instead of the fluxes of a spectral line is twofold: firstly, it allows us to include galaxies that have very weak or even absent emission lines (passive galaxies) while being insensitive to reddening effects.

\subsection{Dominant photoionization mechanism of mixed activity classes}

The area of the BPT enclosed by the lines of \cite{2003MNRAS.346.1055K} and \cite{2001ApJ...556..121K} contains galaxies whose activity is thought to be a combination of star formation and an AGN component. However, photoionization models with pure AGN \citep{1983ApJ...264..105F,1983ApJ...269L..37H,1984A&A...135..341S} or purely old stellar populations \citep{2008MNRAS.391L..29S} can occupy and cover the entire area between the lines of \cite{2003MNRAS.346.1055K} and \cite{2001ApJ...556..121K}. This suggests that in this region of the BPT diagram, these two sub-populations of composite galaxies (starburst-AGN and passive-starburst, and starburst-passive, see Fig. \ref{fig:comp_bpt}) could coexist. By including the D4000 as an age indicator we can remove this degeneracy. In other words, the inclusion of D4000 in the classification scheme will identify galaxies that host old stellar populations and are located in the locus of composite galaxies. For example, \cite{2010ApJ...719.1191T} suggest that the younger stellar component in star-forming galaxies typically have D4000 $\lesssim$ 1.3.

In the case of LINERs, another complex activity class, we observe that they can also be divided into two distinct subclasses based on their origin of activity. As shown in the  right plot of Fig. \ref{fig:merged_decomp_liner}, LINERs segregate in two well-defined sub-populations. Those with greater similarities to AGN galaxies are positioned near the separation line between AGN and LINER galaxies, as delineated by \cite{2007MNRAS.382.1415S} while a second sub-population is located below the first at lower values of log$_{10}$([\ion{O}{III}] $\lambda$5007/H$\beta$). In Fig. \ref{fig:merged_decomp_liner} (right plot) we also see that there is some overlap between these two subpopulations of LINERs. An important point reinforcing our results is that, while a distinct separation line exists between AGN and LINER galaxies on the BPT diagram, \cite{2003ApJ...583..159H} argues that it lacks absolute physical significance. In this sense, the distribution of AGN galaxies covers a broader range than is typically considered. On the other hand, the remaining LINERs located beneath the cluster of the AGN-predicted LINERs are predicted as passive and the lower values of log$_{10}$([\ion{O}{III}] $\lambda$5007/H$\beta$) suggests a weaker excitation mechanism than the AGN-LINERs, that could be explained by the presence of post-AGB stars. The categorization of LINERs into two sub-populations is further validated by \cite{2021ApJ...922..156A}, which demonstrates that LINERs can be classified as hard and soft based on the ionization field's hardness. In addition,  \cite{2021ApJ...922..156A} shows that these two subcategories of LINERs exhibit similar distributions on a BPT diagram as the ones depicted in the right plot of Fig. \ref{fig:merged_decomp_liner}.

In the left plot of Fig. \ref{fig:merged_decomp_liner}, we observe that certain composite galaxies predicted to be powered by an old stellar component are situated just above the \cite{2003MNRAS.346.1055K} line when projected onto the standard BPT diagram. Initially, this outcome may appear unexpected, but it aligns with growing evidence that an increasing contribution from older, hot-evolved stellar populations results in higher [\ion{N}{II}]/H$\alpha$ ratios than typical star-forming galaxies and increasingly lower [\ion{O}{III}]/H$\beta$ rations shifting galaxies towards the lower region of the standard BPT plot \citep{2008MNRAS.391L..29S,2010MNRAS.403.1036C,2017ApJ...840...44B,2019AJ....158....2B}. This trend becomes more evident in Fig. \ref{fig:cp_Byler_bpt}, which displays the distributions of passive-composite, star-forming-composite, and AGN-composite galaxies alongside the trajectory of points from \cite{2019AJ....158....2B}. In this figure, we observe that the increasing age of the stellar populations within a galaxy tends to shift the galaxy's position on a BPT diagram downwards towards lower [\ion{O}{III}]/H$\beta$ ratios, traversing the region of composite galaxies. It is worth noting that AGN-composite galaxies are positioned above the locus of passive-composite galaxies shown here, whereas SF-composite galaxies are closer to the dashed line representing the empirical line of \cite{2003MNRAS.346.1055K} that delineates SF galaxies (left plot of Fig. \ref{fig:merged_decomp_liner}) and closer to the youngest stellar populations as indicated by the colored points.

To further explore these results, we study the specific star-formation rate (sSFR) within our sample of optically selected composite galaxies. The star formation rates and stellar masses are from the galSpecExtra catalog of the SDSS. Figure \ref{fig:comp_ssfr} shows histograms of the sSFR distribution of composite galaxies along with star-forming and passive galaxies. It is evident that composite galaxies displaying spectral characteristics akin to those of SF galaxies (i.e., starburst-composite) tend to have higher sSFR in comparison to those predicted to have similarities with passive galaxies (passive-composite). This observation signifies that our classifier can effectively discern whether the activity within a composite galaxy is attributed to young stars or old stellar populations, thereby aiding in resolving the degeneracy encountered in optical diagnostics \citep{2001ApJ...556..121K,2003MNRAS.346.1055K}.

In conclusion, this new diagnostic tool can be employed not only for the classification of galaxies but also for characterizing the underlying activity of mixed-activity galaxies. However, it should be noted that the probabilities derived from the random forest classifier do not correspond to the contribution of each of the corresponding principal activity mechanisms to the observed spectrum; instead, they represent the likelihood of similarity with each of these classes.

\subsection{Ambiguous classifications} \label{dnc}

\begin{figure}[h]
\begin{center}
\includegraphics[width=8.8cm,height=13cm]{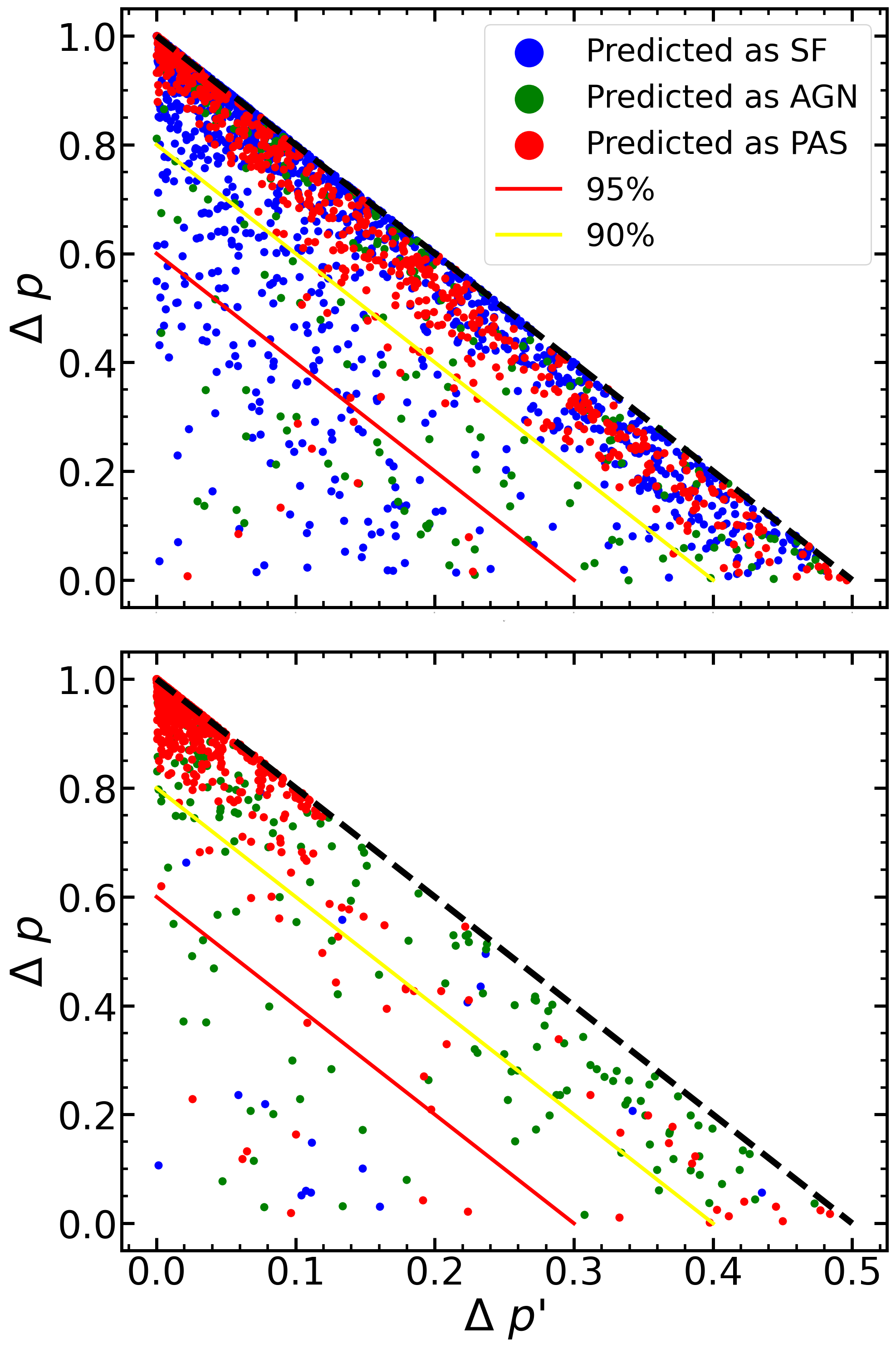}
\end{center}
\caption{$\Delta p$ against $\Delta p'$. In both plots, each dot represents a galaxy of mixed-activity class. The top plot is populated by the composite galaxies, while the bottom plot is populated by the LINER galaxies. In both plots, we observe that the lower-left corner contains objects with $\Delta p$ = $\Delta p'$ = 0, indicating equal probabilities of belonging to all classes, resulting in unreliable classification. The black dashed line represents the extreme reliability line ($\Delta p$ = -2$\cdot \Delta p'$ + 1). The region between the back line and the yellow solid line ($\Delta p$ = -2$\cdot \Delta p'$ + 0.8) contains 90\% of the mixed-class objects for both composites and LINERs, while the region between the black and the solid red line ($\Delta p$ = -2$\cdot \Delta p'$ + 0.6) encloses 95\% of the mixed-class objects for both composites and LINERs.}
\label{fig:dp_dp_P}
\end{figure}

As a general guideline to decide whether objects with maximum predicted probability (max\_$p_{i}$) < 90\% are dominated by two activity mechanisms, we propose selection criteria based on each galaxy to belong in a given class. To establish these, we analyze the differences among their highest ($p_{1}$), second highest ($p_{2}$), and lowest predicted ($p_{3}$) probabilities. Consequently, for the mixed-activity classes of composite and LINER galaxies, we compute the $\Delta p = p_{1}-p_{2}$ and the $\Delta p'=p_{2}-p_{3}$, indicating the probability difference between the first and second class and the probability difference between the second and third class respectively. Figure \ref{fig:dp_dp_P} shows $\Delta p$ plotted against $\Delta p'$ for the sub-sample of composite (top) and LINER (bottom) galaxies. Here, we focus solely on the probability difference as a metric of the discriminating power of this method, rather than the actual classes.

Plotting $\Delta p$ against $\Delta p'$ reveals that the lower-left corner of this plot is populated by objects with similar $\Delta p$ and $\Delta p'$ values. A pure activity galaxy would be situated in the upper left corner ($max\_p_{i}\to 1, \Delta p \to 1, \Delta p' \to 0 $). On the contrary, mixed-class objects tend to be scattered below this ridge. Additionally, we observe that no object crosses the line $\Delta p = -2 \cdot \Delta p' + 1$. This outcome is a direct consequence of having three classes, as the sum of these three predicted probabilities must equal 1 for each object. This line serves as the upper limit for an object in the $\Delta p$-$\Delta p'$ space and can be considered as the extreme reliability line. These two-class objects are situated below  the upper left corner along the $\Delta p  + \Delta p' =1$ ridge and up to the middle of the ridge.  Objects located below the middle of the ridge or below the line and towards the bottom left corner, have increasingly similar probabilities of belonging to all three classes and hence their classifications are not reliable. 

In order to determine the reliability threshold, we identify the area in the $\Delta p$-$\Delta p'$ diagram (Fig. \ref{fig:dp_dp_P}) that includes 90\% of the galaxies below the maximum probability line. This process establishes a reliability threshold that encompasses 90\% of each population, similar to the reliability thresholds set for the pure classes (see Table \ref{tab:tab_nwcldef} in Sect. \ref{res}). To define this reliability criterion, we initially use the inequality $\Delta p < m \cdot \Delta p' + C$, where $m$ and $C$ are constants. We set $m= -2$, which is the slope of the maximum reliability line (represented by the black dashed lines in Fig. \ref{fig:dp_dp_P}), and then shift this line towards the lower left corner until 90\% of the objects are enclosed, in order to find the value of $C$. Based on our findings, the equation that meets these criteria is $\Delta p < -2 \cdot \Delta p' + 0.8$. Therefore, any object adhering to the relationship $\Delta p < -2 \cdot \Delta p' + 0.8$ is designated as having an inconclusive classification. It's worth noting that the value of $C$ can be adjusted based on user preference. An extreme class of inconclusive galaxies are those that have similar probabilities between all three classes.

In Sect. \ref{sec44} and \ref{sec45}, it was found that for a small subset of galaxies, the classification probabilities are equally split, i.e., the probability of an object to belong in an activity class is the same or very similar across all classes and as a result we cannot characterize the dominant activity mechanism or the two most prevailing ones for these galaxies. As anticipated, this issue primarily arises in the mixed activity classes, specifically for galaxies spectroscopically identified as composites and LINERs.

\subsection{Comparison with other methods}

The use of equivalent width as an activity diagnostic has been explored in the past. \cite{2010MNRAS.403.1036C} considered diagnostic diagrams involving the EW of H$\alpha$ and the [\ion{N}{II}]/H$\alpha$ line ratio (WHAN diagrams) in order to include objects with weak emission lines which are typically excluded from most diagnostic tools. This led to a three-class scheme consisting of star-forming, AGN, and weak emission-line objects. The latter category comprises objects characterized by low S/N (below 3) in the emission lines conventionally employed in the BPT diagram, except for the emission lines of [\ion{N}{II}] and H$\alpha$. The classification of this particular activity class bears resemblance to the categorization applied to our definition of passive galaxies, as explained in Sect. \ref{sec:23}. This similarity emerges in the transition of galaxies exhibiting weak emission lines stemming from excitation by old stellar populations to a state of complete passivity, indicative of a dormant system marked solely by absorption lines. In this respect, the definition of passive galaxies presented in Sect. \ref{sec:23}, which is based on photometric criteria, is more general since it encompasses completely passive galaxies that do not exhibit any Balmer or forbidden lines.

To evaluate the performance of our diagnostic tool against the WHAN diagnostic, we consider all galaxies that meet the eligibility criteria defined for our sample in Sect. \ref{sec24}, with the additional condition that the S/N of both [\ion{N}{II}] and H$\alpha$ is above 3, as required by the WHAN diagnostic. Subsequently, we consider the spectroscopic classification (SoDDA classifier) as the ground truth for star-forming and AGN galaxies, and the photometric selection as the ground truth for passive galaxies.
\begin{table}[h!]
\centering
\caption{Comparison of the activity classification between the DONHa and the WHAN diagnostic \citep{2010MNRAS.403.1036C}.}
\begin{tabular}{ccccccc}
\multicolumn{1}{l}{}               & 
\multicolumn{4}{c}{SoDDA-SFG}                               \\ \hline
\multicolumn{6}{c}{WHAN}  \\ %\hline
\multirow{8}{*}{\rotcell{DONHa}} %
& & SFG & AGN & Passive  
 & Total  \\ \hline
 & SFG   & 10450 & 318 & 0 & 10768 \\
 & AGN   & 34 & 0 & 1 & 35 \\
 & Passive   & 14 & 4 & 1 & 19 \\
 & Total & 10498 & 322 & 2 &  &   
\\ \hline \hline
\\
\multicolumn{6}{c}{SoDDA-SFG}                               \\ \hline
\multicolumn{6}{c}{WHAN}                               \\ %\hline
\multirow{8}{*}{\rotcell{DONHa}} 
& & SFG & AGN & Passive  
 & Total  \\ \hline
 & SFG     & 5 & 1 & 0 & 6 \\
 & AGN     & 25 & 289 & 64 & 378 \\
 & Passive     & 0 & 0 & 0 & 0 \\
 & Total   & 30 & 290 & 64 &  &  
 \\
 \hline \hline
\\
\multicolumn{6}{c}{Passive}                               \\ \hline
\multicolumn{6}{c}{WHAN}                               \\ %\hline
\multirow{8}{*}{\rotcell{DONHa}} %  
& & SFG & AGN & Passive 
 & Total  \\ \hline
 & SFG     & 13 & 7 & 0 & 20 \\
 & AGN     & 0 & 8 & 29 & 37 \\
 & Passive     & 24 & 37 & 6281 & 6342 \\
 & Total   & 37 & 52 & 6310 &  &  
\\ \hline \hline
\end{tabular}
\label{tab:wha}
\tablefoot{For this comparison, we use the sample of SF, AGN, and passive galaxies which we consider as the ground truth (SoDDA classification, for SF and AGN) and the photometric selection for passive galaxies (Sect. \ref{datasample}).}
\end{table}
Table \ref{tab:wha} is segmented into three distinct subtables (indicated in the title of each subtable) and serve as the ground truth for their classification. Within each subtable the rows correspond to the classification from our disgnostic, while columns correspond to objects classified by the WHAN diagnostic.

In Table \ref{tab:wha} we see that our diagnostic performs better as it manages to identify a larger fraction of the objects within each activity class. Notably, for the AGN class, our diagnostic accurately identifies approximately 99\% of true AGN, whereas the WHAN diagnostic correctly identifies only around 76\%. Another key advantage of our diagnostic is its capability to include galaxies that exhibit absorption features, and not only to those with emission lines. This feature enhances the overall efficacy of our diagnostic and broadens its applicability to diverse datasets.

Another interesting fact from Table \ref{tab:wha} is that there are 318 objects that the DONHa diagnostic has correctly identified as star-forming (according to the ground truth based on the SoDDA classifications), which were classified as AGN based on the WHAN diagnostic. After further investigation, we found that these objects have -0.8 < [\ion{O}{III}]/H$\beta$ < -0.2 and -0.32 < [\ion{N}{II}]/H$\alpha$ < -0.18, placing them low at the rightmost tip of the star-forming galaxy locus on a BPT diagram (close to the line of composite galaxies defined by \cite{2003MNRAS.346.1055K}). These galaxies have equivalent widths of H$\alpha$ > 6 $\AA$, which classifies them as AGN in the WHAN diagnostic. However, their [\ion{O}{III}]/H$\beta$ ratio is not strong enough to characterize them as AGN in the two-dimensional BPT diagram or the four-dimensional SoDDA diagnostic. In addition, their [\ion{N}{II}]/H$\alpha$ ratios indicate relatively high metallicities. After visual inspection of their optical SDSS images, we see that these galaxies have prominent bulges, an indication of the presence of old stellar populations \citep{2000A&A...355..949F,2001A&A...376..878O}. Also, their optical spectra show high H$\alpha$ and [\ion{N}{II}] but low [\ion{O}{III}] fluxes, as well as absorption features around H$\beta$ which is another characteristic of evolved stellar populations.

% Concluding this analysis, it is noteworthy to mention that in the study conducted by \cite{2010MNRAS.403.1036C}, the authors argue that the region between the boundary lines established by \cite{2003MNRAS.346.1055K} and \cite{2001ApJ...556..121K} in the BPT diagram, commonly referred to as the "mixing sequence," can encompass galaxies whose activity is not solely driven by a combination of starburst and AGN components. It is also worth noting that this finding aligns with the results obtained in our analysis, as illustrated in Fig. \ref{fig:merged_decomp_liner} (left plot).

\section{Conclusions} \label{concl}

In this study, we have undertaken a classification task centered around the identification of the three primary excitation mechanisms governing galaxy activity: star formation, AGN activity, and photoionization by old stellar populations. Furthermore, we have delved into the classification of galaxies exhibiting mixed-activity profiles, specifically composite and LINER galaxies. Our investigation has demonstrated the feasibility of characterizing these mixed-activity galaxies by determining their most probable source of activity among the three principal mechanisms. Furthermore, we introduce a new two-dimensional diagnostic diagram, the DO3, that encompasses all forms of galaxy activity. This diagnostic is highly versatile and applicable to a wide range of data sets, based on its simplicity and narrow spectral range. Below, we provide a summary of the results and conclusions derived from this work.

\begin{enumerate}
  \item We present a galaxy activity diagnostic tool utilizing machine-learning methods, relying solely on four spectral features (the equivalent widths of the H$\alpha$, [\ion{O}{III}] $\lambda$5007, [\ion{N}{II}] $\lambda$6584 lines, and the D4000 index), that allows us to seamlessly integrate excitation by  star formation, AGN as well as old stellar populations, with excellent reliability and completeness outperforming the currently available methods.
  \item We introduce a novel two-dimensional activity diagnostic diagram (the DO3 diagram), involving the D4000 break against log$_{10}$([EW(\ion{O}{III} $\lambda$5007)]$^{2}$) line. To the best of our knowledge, this represents the first two-dimensional diagnostic plot that encompasses all activity classes, while providing insights into the primary ionization The requirement for only a brief observed rest-frame spectral range for our DO3 diagnostic greatly extends its applicability to high-redshift galaxies observed in wide area spectroscopic campaigns or even more distant galaxies observed with JWST.
  \item The application of our diagnostic method on the composite galaxies identifies two types of composite galaxies: those powered by AGN alongside star forming activity, and those powered by star formation as well as older stellar populations. The use of the D4000 break allows us to discriminate between these two types.
  \item Concerning LINER galaxies, our analysis enables discrimination between those dominated by AGN activity and those excited by old stellar populations. Our analysis suggests that the ionization mechanism of 72\% of all BPT LINERS can be solely attributed to evolved-stellar populations and not AGN activity.
  \item The probabilities generated by our diagnostic can serve as an indicator for characterizing the principal activity mechanism of the host galaxy.
\end{enumerate}

\section{Data availability}
The code, along with detailed application instructions, can be found in this GitHub repository\footnote[2]{\url{https://github.com/BabisDaoutis/DONHaClassifier}}.

\begin{acknowledgements}
We sincerely thank the anonymous referee for their valuable comments and suggestions, which greatly enhanced the quality and clarity of this manuscript. We would like to express our sincere gratitude to prof. Samir Salim for generously providing GALEX UV photometry that was essential for this research.
CD and EK acknowledge support from the Public Investments Program through a Matching Funds grant to
the IA-FORTH. KK is supported by the project ''Support of the international collaboration in astronomy (Asu mobility)'' with the number: CZ 02.2.69/0.0/0.0/18\_053/0016972. Funding for the Sloan Digital Sky 
Survey IV has been provided by the 
Alfred P. Sloan Foundation, the U.S. 
Department of Energy Office of 
Science, and the Participating 
Institutions. 

SDSS-IV acknowledges support and 
resources from the Center for High 
Performance Computing  at the 
University of Utah. The SDSS 
website is www.sdss.org.

SDSS-IV is managed by the 
Astrophysical Research Consortium 
for the Participating Institutions 
of the SDSS Collaboration including 
the Brazilian Participation Group, 
the Carnegie Institution for Science, 
Carnegie Mellon University, Center for 
Astrophysics | Harvard \& 
Smithsonian, the Chilean Participation 
Group, the French Participation Group, 
Instituto de Astrof\'isica de 
Canarias, The Johns Hopkins 
University, Kavli Institute for the 
Physics and Mathematics of the 
Universe (IPMU) / University of 
Tokyo, the Korean Participation Group, 
Lawrence Berkeley National Laboratory, 
Leibniz Institut f\"ur Astrophysik 
Potsdam (AIP),  Max-Planck-Institut 
f\"ur Astronomie (MPIA Heidelberg), 
Max-Planck-Institut f\"ur 
Astrophysik (MPA Garching), 
Max-Planck-Institut f\"ur 
Extraterrestrische Physik (MPE), 
National Astronomical Observatories of 
China, New Mexico State University, 
New York University, University of 
Notre Dame, Observat\'ario 
Nacional / MCTI, The Ohio State 
University, Pennsylvania State 
University, Shanghai 
Astronomical Observatory, United 
Kingdom Participation Group, 
Universidad Nacional Aut\'onoma 
de M\'exico, University of Arizona, 
University of Colorado Boulder, 
University of Oxford, University of 
Portsmouth, University of Utah, 
University of Virginia, University 
of Washington, University of 
Wisconsin, Vanderbilt University, 
and Yale University.
\end{acknowledgements}

\bibliographystyle{aa}
\bibliography{references}

\begin{appendix} 

\section{Algorithm optimization} \label{appendix1}

Hyperparameters are parameters that enable an algorithm to be tailored to the specific needs of a given problem. Not all hyperparameters are always relevant and require optimization; some perform effectively with their default settings, while others can significantly impact the performance of the algorithm even with minor adjustments. Upon investigation, we have determined that the performance of the random forest algorithm for our problem is primarily influenced by the following hyperparameters:
\begin{itemize}
    \item[$-$] \texttt{n\_estimators} : The number of trees.
    \item[$-$] \texttt{max\_depth} : The maximum depth of a each tree.
    \item[$-$] \texttt{min\_samples\_split} : The minimum number of samples required for node splitting
    \item[$-$] \texttt{min\_samples\_leaf} : The minimum number of samples in a leaf node.
    \item[$-$] \texttt{max\_leaf\_nodes} : The maximum number of final nodes in a tree.
    \item[$-$] \texttt{max\_samples} : The number of samples from the training set need to build a tree.
    \item[$-$] \texttt{criterion} : Measure of the quality of a split.
\end{itemize}

Further investigation confirms that the remaining hyperparameters do not significantly affect the performance, and as such, they are left at their default values as imported with the \texttt{RandomForestClassifier}.

\begin{figure}[h]
\begin{center}
\includegraphics[scale=0.25]{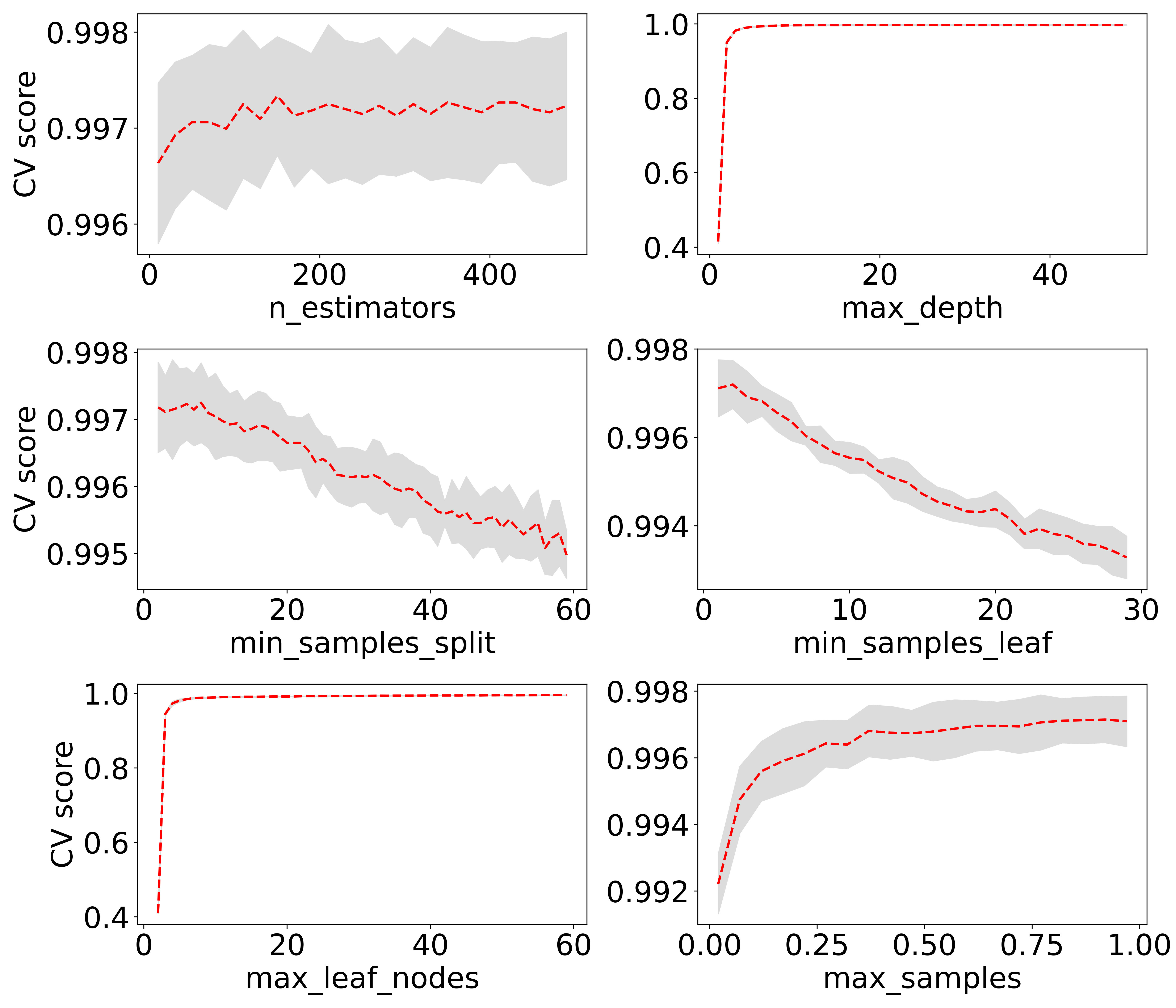}
\end{center}
\caption{Accuracy scores (CV scores) against all important random forest hyperparameters. CV scores have been computed through cross-validation, varying a single random forest hyperparameter while keeping all others constant. The red dashed line represents the average of these scores, and the gray shaded area indicates their standard deviation (1$\sigma$). By analyzing these curves, we can determine the range for each hyperparameter, which is crucial for reducing the number of hyperparameter combinations during the grid-search optimization procedure.}
\label{fig:val_curves}
\end{figure}

\begin{table}[h]
\caption{Heperparameter search ranges and optimal values.}
\centering
\begin{tabular}{l c c c c }
\hline\hline
Parameter & Search range & Best value \\
\hline
\texttt{n\_estimators} & 100-200 & 160 \\
\texttt{max\_depth} & - &  \texttt{'default'}\\
\texttt{min\_samples\_split} & 25-40 & 38 \\
\texttt{min\_samples\_leaf} & 2-20 & 7 \\
\texttt{max\_leaf\_nodes} & - & \texttt{'default'} \\
\texttt{max\_samples} & 0.1-1.0 & 1.0 \\
\texttt{class\_weight} & - &  \texttt{'balanced'} \\
\texttt{criterion} & - & ' \texttt{Gini'} \\
\hline
\label{hyp}
\end{tabular}
\tablefoot{The values presented in the third column are the ones we adopted for the implementation of the random forest algorithm in this project.}
\end{table}

The conventional approach for determining the optimal values for the relevant hyperparametrs involves assessing the performance of the random forest across a multidimensional grid of their values. Given the multitude of hyperparameters and the extensive range of potential values, we employ validation curves to reduce the number of grid points and render the problem more manageable. These curves depict how a selected performance score, in our case accuracy, changes as a function of the various values that a specific hyperparameter can take. To elaborate, our approach to explore the hyperparameter ranges is as follows: for each hyperparameter of interest, we vary it across its feasible range of values while maintaining all other hyperparameters at their default values. Subsequently, we record the corresponding performance score (the balanced accuracy in our case) for the entire range of values explored. We repeat this procedure for each of the other hyperparameters. In Figure \ref{fig:val_curves}, we present the validation curves for all significant hyperparameters that influence the performance of the classifier.

To evaluate the performance of the algorithm and select optimal values for each hyperparameter, we use k-fold cross-validation method. In this process, the training data are divided into k subsets, with one used for testing while the rest train the algorithm. The algorithm is trained K times, each time using a different subset as the test set. The reported scores are averages across these cycles, providing not only a performance assessment but also an estimation of score uncertainty. The algorithm utilized for the optimization task is \texttt{GridSearchCV}, from the \texttt{scikit-learn} library. After analyzing the validation curves, we determined the optimal ranges and employed them as inputs for the \texttt{GridSearchCV} algorithm to construct a multidimensional search grid. The ranges and the optimal hyperparameter values are presented in Table \ref{hyp}.

% These curves depict how a selected performance score, in our case accuracy, changes as a function of the various values that a specific hyperparameter can assume. To elaborate, our approach to explore the hyperparameter ranges is as follows: for each hyperparameter of interest, we vary it across its feasible range of values while maintaining all other hyperparameters at their default settings. Subsequently, we record the corresponding performance score (the balanced accuracy in our case) for the entire range of values explored. We repeat this procedure for each of the other hyperparameters.

\section{Separating boundaries between activity classes on the two-dimensional activity diagram} \label{appendix2}

Here we present the selection criteria defined for the DO3 activity diagnostic diagram as described in Sect. \ref{sec45}. These equations indicate the boundaries that delineate the activity classes defined on Fig. \ref{fig:neo_BPT}. 
In the following equations we set $y \equiv$ D4000 and $x \equiv$ 2 log$_{10}$(|EW(\ion{O}{III} $\lambda$5007)|).\\

\noindent Pure star-forming:
\begin{equation}
\begin{aligned}
    y < -0.003~x^{4} + 0.034~x^{3} - 0.132~x^{2} + 0.090~x^{2}+ 1.411
\end{aligned}
\end{equation}
Pure AGN:
\begin{equation}
y < 1.476~x - 0.006
\end{equation}

and 
\begin{equation}
    y > \frac{0.597}{x-0.081} + 1.068
\end{equation}
% or instead of (3),
% \begin{equation}
%     \text{D4000} > -0.12~x + 1.58
% \end{equation}
Pure passive:
\begin{equation}
    x < 0.90
\end{equation}

and 
\begin{equation}
    y > 0.119~x + 1.511
\end{equation}
Starburst-passive:
\begin{equation}
    x < 0.90
\end{equation}

and
\begin{equation}
\begin{aligned}
    y > -0.003~x^{4} + 0.034~x^{3} - 0.132~x^{2} + 0.090~x+ 1.411
\end{aligned}
\end{equation}

and
\begin{equation}
    y < 0.119~x + 1.511
\end{equation}
Starburst-AGN:
\begin{equation}
\begin{aligned}
    y > -0.003~x^{4} + 0.034~x^{3} - 0.132~x^{2} + 0.090~x + 1.411
\end{aligned}
\end{equation}

and
\begin{equation}
y < 1.476~x - 0.006
\end{equation}

and
\begin{equation}
    y < \frac{0.597}{x-0.081} + 1.068
\end{equation}
AGN-passive:
\begin{equation}
% \begin{equation}
\begin{aligned}
    y > -0.003~x^{4} + 0.034~x^{3} - 0.132~x^{2} + 0.090~x + 1.411
\end{aligned}
% \end{equation}
\end{equation}

and 
\begin{equation}
y > 1.476~x - 0.006
\end{equation}

and
\begin{equation}
    x > 0.90
\end{equation}
% where $y \equiv$ D4000 and the $x \equiv$ log$_{10}$([EW(\ion{O}{III} $\lambda$5007)]$^{2}$).

\end{appendix}
\end{document}